\documentclass[onecolumn]{aastex631}

\usepackage{xspace}
\usepackage{xcolor}
\usepackage{graphicx}
\usepackage{amsmath,bm}

\usepackage{todonotes}

\newcommand{\pt}{\ensuremath{p_\mathrm{T}^{(0)}}\xspace}
\newcommand{\mpt}{\ensuremath{\langle\pt\rangle}}
\newcommand{\vdir}{$v_1$\xspace}
\newcommand{\vel}{$v_2$\xspace}
\newcommand{\ekin}{$E_{\mathrm{kin}}/A$\xspace}
\newcommand{\rap}{$y^{(0)}$\xspace}
\newcommand{\erat}{$E_{\rm rat}$\xspace}
\newcommand{\agev}{$A$\,GeV\xspace}
\newcommand{\chisqrt}{$\chi^2/n.d.f.$\xspace}

\begin{document}

\title{Flow and Equation of State of nuclear matter at $\bm{E_{\rm kin}/A=0.25}$-1.5~GeV with the SMASH transport approach}

\author[0000-0001-5086-8658]{L.A. Tarasovi\v{c}ov\'{a}}
\affiliation{Pavol Jozef \v{S}af\'arik University, \v{S}rob\'arova 2, 04011, Ko\v{s}ice, Slovakia}

\author[0000-0001-8437-0946]{J. Mohs}
\affiliation{Frankfurt Institute for Advanced Studies, Frankfurt am Main, Germany}
\affiliation{Institute for Theoretical Physics, Goethe University, Frankfurt am Main, Germany}

\author[0000-0002-2372-6117]{A. Andronic}
\affiliation{Institut f\"ur Kernphysik, Universit\"at M\"unster, Germany}

\author[0000-0002-6213-3613]{H. Elfner}
\affiliation{GSI Helmholtzzentrum f\"ur Schwerionenforschung, Planckstr. 1, 64291 Darmstadt, Germany}
\affiliation{Frankfurt Institute for Advanced Studies, Frankfurt am Main, Germany}
\affiliation{Institute for Theoretical Physics, Goethe University, Frankfurt am Main, Germany}
\affiliation{Helmholtz Research Academy Hesse for FAIR (HFHF), GSI Helmholtz Center, Campus Frankfurt, Max-von-Laue-Straße 12, 60438 Frankfurt am Main, Germany}

\author[0000-0002-2805-0195]{K.-H. Kampert}
\affiliation{University Wuppertal, Department of Physics, 42117 Wuppertal, Germany}

\date{Received: date / Accepted: date}

\begin{abstract}

We present a comparison of directed and elliptic flow data by the FOPI collaboration in Au--Au, Xe--CsI, and Ni--Ni collisions at beam kinetic energies from 0.25 to 1.5 GeV per nucleon to simulations using the SMASH hadronic transport model. 
The Equation of State is parameterized as a~function of nuclear density and momentum dependent potentials are newly introduced in SMASH. 
With a~statistical analysis, we show that  within the present status of the
SMASH transport model, the collective flow data at lower energies is in the~best agreement with a~soft momentum dependent potential, while the~elliptic flow at higher energies requires a~harder momentum dependent Equation of State. 

\end{abstract}



\section{Introduction} \label{sec:intro}

The Equation of State (EoS) of nuclear matter at densities a few times the normal nuclear matter density has recently attracted increased attention because it influences the properties of neutron stars and neutron star mergers, with the~latter now being probed by gravitational wave interferometers, see e.g.~\cite{Oechslin:2006uk,Fattoyev:2020cws}. 
Independent constraints of the~EoS are provided by laboratory experiments of heavy-ion collisions performed at beam kinetic energies in the~range of $E_{\rm kin} \sim 0.1$ to a few GeV per nucleon (\agev) in the~laboratory frame~\cite{Andronic:2006ra,Sorensen:2023zkk,MUSES:2023hyz}.
Through a comparison of the~measured collective flow data and transport model calculations, a range of constrains were achieved in the~last decades, see e.g.~\cite{Danielewicz:2002pu,Oliinychenko:2022uvy,OmanaKuttan:2022aml,Russotto:2023ari}.
Further information about the EoS from heavy-ion collisions were extracted using the production of strange mesons below threshold~\cite{Fuchs:2000kp}.  
Moreover, it has recently been shown that by combining data from astrophysical multi-messenger observations and flow measurements in heavy-ion collisions within a Bayesian analysis of thousands of nuclear theory motivated EoS versions~\cite{Huth:2021bsp}, further important constraints on the~EoS can be achieved. 
This can start a new multidisciplinary field of science.

Nevertheless, the uncertainties in the EoS in the range of densities $\rho_B/\rho_0 \simeq$ 1-3 (with $\rho_0$ being the~normal nuclear matter density) remain large up to this day~\cite{OmanaKuttan:2022aml,Lynch:2021xkq} and are to some extent model dependent.
As a simple parametrization of the EoS may not be able to describe consistently the flow data across a broad range of beam energies, Bayesian approaches were recently successfully employed, leading to indications of a softening of the EoS at densities $\rho_B/\rho_0 \simeq$ 3-4~\cite{Oliinychenko:2022uvy,OmanaKuttan:2022aml}. 
In~\cite{LeFevre:2015paj, Cozma:2024cwc}, it has been shown, that the potential has to be dependent on momentum to achieve a reasonable description of collective flow data. 
The variability in the transport model approaches and also the constraints on symmetric matter EoS from flow (and other observables) in heavy-ion collisions (see e.g.~\cite{FOPI:2004bfz}) led to the Transport Model Evaluation Project (TMEP)~\cite{TMEP:2022xjg}. 
The SMASH transport model~\cite{SMASH:2016zqf}, which we use in the present paper, is part of the TMEP.

In this paper, we compare model calculations with FOPI data on the directed and 
elliptic flow coefficients $v_1$~\cite{FOPI:2003fyz} and $v_2$~\cite{FOPI:2004bfz}, respectively, spanning beam energies from 0.25 to 1.5~\agev.
SMASH has been already employed for the~description of HADES data at 1.2~\agev~\cite{Mohs:2020awg} and in the multi-GeV range for the description of the STAR BES data~\cite{Oliinychenko:2022uvy}, but its application to the beam energies down to 0.25~\agev is studied here for the first time.

\newpage
\section{Model description}

The transport model SMASH \cite{SMASH:2016zqf} is a Boltzmann-Uehling-Uhlenbeck-type model with open-source code\footnote{Wergieluk, A. (2024) ‘smash-transport/smash: SMASH-3.1’. Zenodo. doi: 10.5281/zenodo.10707746.}, used over a wide range of collision energies, either standalone~\cite{Mohs:2020awg,Oliinychenko:2022uvy} or as an afterburner for hydrodynamic calculations~\cite{Schafer:2021csj}.
A~large set of 208 stable hadrons and resonances are included in the model and Pauli-blocking is taken into account.
For the description of resonances in the model, vacuum Breit-Wigner spectral functions are assumed and no in-medium modifications of the cross section is implemented.
We refer the reader to~\cite{SMASH:2016zqf} for further details on the model in general.
As nuclear potentials can be related to the EoS, which plays a major role for the description of a heavy-ion collision in the energy range considered here, we will first describe the ones employed in this work.

\subsection{Potentials}

We incorporate a Skyrme and a symmetry potential of the form
\begin{equation}
    U_{\mathrm{Skyrme}} = A \left(\frac{\rho_B}{\rho_0} \right) + B\left( \frac{\rho_B}{\rho_0}\right)^\tau
    \label{eq:skyrme}
\end{equation}
\begin{equation}
    U_\mathrm{symmetry} = \pm 2S_\mathrm{pot} \frac{\rho_{I3}}{\rho_0}\,,
    \label{eq:symmetry}
\end{equation}
where $\rho_B$ denotes the baryon density, $\rho_0$ the saturation density and $A$, $B$ an $\tau$ are parameters as given in Tab.~\ref{tab:eos_summary}.
The~sign in the~symmetry potential depends on the sign of the isospin of the considered particle. $\rho_{I3}$ denotes the~density of the relative isospin projection $I_3/I$ and $S_\mathrm{pot}$ is a constant which is fixed to $18\,\mathrm{MeV}$ as agreed in the  code comparison effort~\cite{TMEP:2016tup}.
We further add a~momentum-dependent term to the potential for which we use the form suggested in Ref.~\cite{Welke:1988zz}
\begin{equation}
U_\mathrm{MD}(\rho,\mathbf{p})=\frac{2C}{\rho_0}g\int\frac{d^3p'}{(2\pi\hbar)^3}\frac{f(\mathbf{r}, \mathbf{p}')}{1+\left(\frac{\mathbf{p}-\mathbf{p}'}{\hbar\Lambda}\right)^2} \,.
\label{eq:momentum_dependence}
\end{equation}
Here, $\mathbf{p}$ denotes the momentum three-vector of the particle of interest, $g$ is the degeneracy factor and $C$ and $\Lambda$ are parameters, see Tab.~\ref{tab:eos_summary}.
In order to simplify the integral, we follow the implementation in GiBUU~\cite{Buss:2011mx} and apply a~cold nuclear matter approximation for the distribution function $f(\mathbf{r},\mathbf{p}) = \Theta(p_F(\rho_B(\mathbf{r}))-|\mathbf{p}|)$ so that the integral can be solved analytically.
Consistent with this approximation, the degeneracy factor is $g=4$ due to spin and isospin of nucleons.
Even though the cold nuclear matter assumption is made, the (momentum-dependent) potential is also applied for all baryonic resonances.
For strange hadrons and resonances, the force is scaled with the fraction of strange quarks contained in the hadron or resonance.

\begin{table}[b]
\begin{center}
\begin{tabular}{|c|c|c|c|c|c|c|} 
\hline
  & \multicolumn{2}{c|}{Hard EoS} & \multicolumn{2}{c|} 
  {Medium EoS} & \multicolumn{2}{c|}{Soft EoS} \\ \cline{2-7}
  Potentials & Const. (H) & Mom. dep. (HM) & Const.  & Mom. dep. (MM) & Const.  & Mom. dep. (SM) \\ \hline 
  $A\,[\mathrm{MeV}]$ & -124.4 & -11.13  &-- & -29.3& -- &-108.6 \\ 
 $B\,[\mathrm{MeV}]$ & 71.0 & 38.28 & -- &57.2 & -- &136.8\\ 
 $\tau$ & 2.0 & 2.4 & -- & 1.76 & -- &1.26\\
 $C\,[\mathrm{MeV}]$ & -- & -63.12 & -- & -63.6& -- &-63.6\\
 $\Lambda \,[ $c$/\mathrm{fm}]$ & -- & 2.12 & -- & 2.13& -- &2.13\\
 $\kappa \,[\mathrm{MeV}]$ & 380 & 380 & -- &290 & -- & 215 \\
\hline
\end{tabular}
\caption{Summary of the parameters for the versions of the EoS employed in this work.}
\label{tab:eos_summary}

\end{center}
\end{table}

As the potentials are not written in a covariant form, one has to fix the frame in which they are evaluated for a~relativistic description of the system.
We therefore calculate the single particle potentials using the above equations only in the local rest frame (LRF) and find the energy in the calculation frame making use of the invariance of $p_\mu p^\mu$ and solving
\begin{equation}
    E_\mathrm{calc}^2 - c^2 p_\mathrm{calc}^2 = E_\mathrm{LRF}^2- c^2 p_\mathrm{LRF}^2
\end{equation}
for $E_\mathrm{calc}$, which is the energy in the calculation frame for a given momentum $\mathbf{p}_\mathrm{calc}$ in that frame.
Note that since $\mathbf{p}_\mathrm{LRF}$ is obtained by boosting the four-momentum from the calculation frame to the local rest frame and $E_\mathrm{LRF}~=~\sqrt{m^2 c^4 + c^2 p_\mathrm{LRF}^2}+U(\rho,p)$, the equation needs to be solved numerically.
The gradient of the energy in the calculation frame is obtained using finite difference with a lattice spacing of 1 fm and the momenta are updated after each time step of 0.1 fm/$c$ according to 
\begin{equation}
    \dot{\mathbf{p}}_\mathrm{calc} = -\nabla E_\mathrm{calc}\,.
\end{equation}
The parameters of the potentials $A$, $B$, $\tau$, $C$ and $\Lambda$ are given in Table \ref{tab:eos_summary} and are fixed for a given incompressibility $\kappa$ to reproduce nuclear ground state properties and the optical potential~\cite{Cooper:2009zza}.
The potential should also include an~electromagnetic part but it is neglected in this work as it is numerically quite expensive in a BUU-type calculation due to the long range of the interaction.
A further shortcoming of the model is that the influence of the potentials are not considered in the collision term.
Including the potential for enforcing the energy conservation in the collision term and changing the thresholds accordingly has been studied in~\cite{Zhang:2017nck, Cozma:2014yna} but are neglected in the present work.

For the evaluation of the nuclear potentials, the density needs to be calculated.
A Lorentz-contracted Gaussian smearing kernel is applied in order to obtain a smooth density profile (see Ref.~\cite{Oliinychenko:2015lva} for details on the smearing kernel).
As described in the following section, we present calculations with coalescence and with dynamic light nuclei formation via scatterings.
The calculations with coalescence are using only one test particle to keep the coalescence simple. In this case, we obtain good statistics for the density calculation using 300 parallel ensembles.
For the dynamic formation of light nuclei, we use 25 test particles.

\begin{figure}[t]
    \centering
    \includegraphics[width=0.5\textwidth]{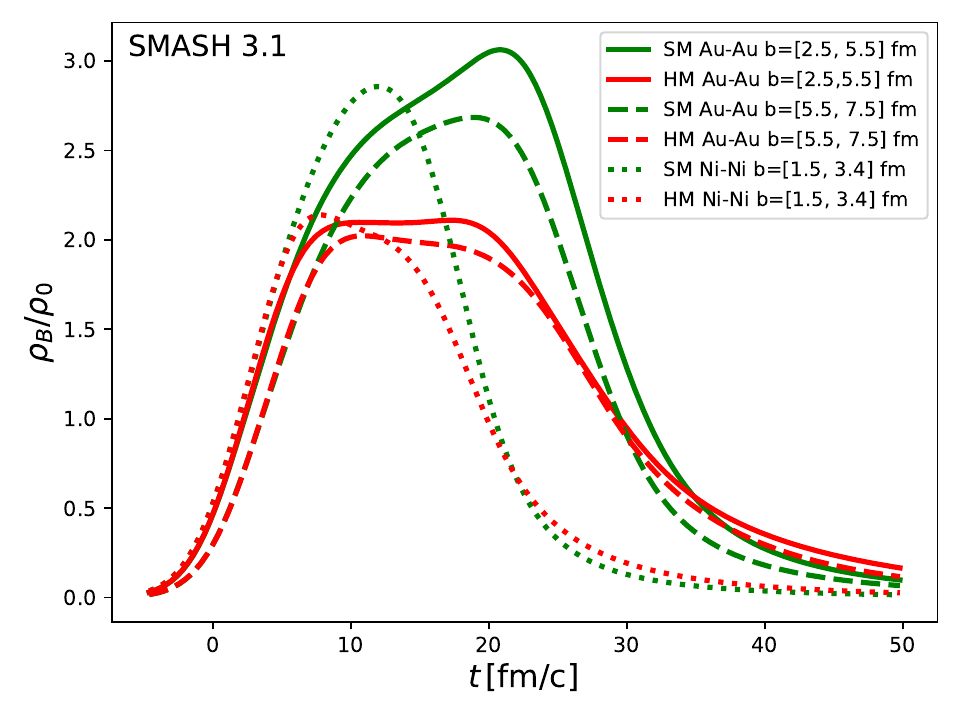}%
    \includegraphics[width=0.5\textwidth]{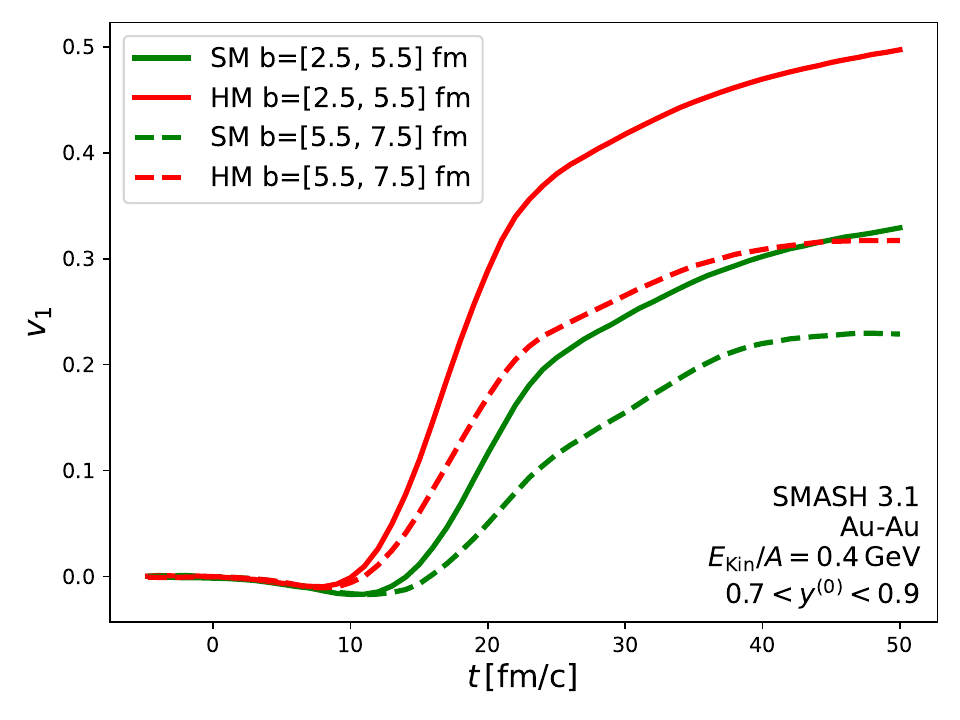}
    \caption{Left: normalized baryon density for two centrality classes in Au--Au collisions and one centrality for Ni--Ni collisions at 0.4 \agev as a function of time. Right: the directed flow \vdir (see below) of protons at forward rapidity for Au--Au collisions in two centrality classes as a function of time.}
    \label{fig:dens_t}
\end{figure}

The collision time evolution of the average density in the central cell of the participant zone at \ekin~=~0.4\,GeV is shown in Fig.~\ref{fig:dens_t} (left) for two impact parameter ranges of Au-Au collisions. The impact parameter range shown for Ni--Ni collisions corresponds to the more central Au--Au class. For both systems, 
soft and hard EoS parametrizations are compared. As expected, a soft EoS leads to larger average densities than the hard EoS. While differences between the two EoS parametrizations are significant, the dependence on the system size is rather weak. The~duration of the~dense phase is shorter in Ni--Ni compared to Au--Au following the~expectation.
In Fig.~\ref{fig:dens_t} (right), the evolution of the~directed flow coefficient $v_1$ (see below) with time for two centralities in Au--Au collisions is shown.
The directed flow is larger for the stiffer EoS.
It mainly builds up between 10 and 30\,fm/$c$ and continues to rise at later times when the density is already significantly smaller in the central cell.
We observe that the directed flow builds up earlier and more rapidly with a stiff EoS, which can be associated to a stronger bounce off compared with the case of soft EoS. Our simulations for the results shown further on extend to 100 fm/$c$.

\subsection{Light nuclei formation}
\label{sec:lnf}

Another important aspect of modeling heavy-ion collisions in the considered energy range is the formation of light nuclei, as a large fraction of nucleons is bound~\cite{FOPI:2010xrt}. 
Light nuclei exhibit a stronger sensitivity to collective flow~(see Ref.~\cite{FOPI:2003fyz} and references therein).
The mechanisms of light nuclei formation are interesting per se.
They were studied in the pBUU transport model~\cite{Danielewicz:1991dh} and are currently investigated over a broad range of collision energies in SMASH~\cite{Oliinychenko:2020znl,Staudenmaier:2021lrg}, within the PHQMD model~\cite{Coci:2023daq,Kireyeu:2024woo}, or in kinetic approaches~\cite{Sun:2022xjr}.

In this work, we mainly apply a coalescence model where light nuclei are identified in the final state of the~calculation~\cite{Mohs:2020awg}.
In order to decide whether a pair of nucleons or nuclei coalesce, we boost to their two-particle center-of-mass~(c.m.s.) frame, find the latest time where one of the two has taken part in a collision, and determine the positions of the candidates at that time.
If the distance of the two in the c.m.s.\ frame is smaller than a threshold of $3\,\mathrm{fm}$ and the momentum difference, also in the c.m.s.\ frame, is below the threshold of $300\,\mathrm{MeV}/c$, coalescence is possible.
In this setup, we calculate the densities for the potentials using 300 parallel ensembles and the light nuclei are identified in each ensemble separately.

We also performed our calculations treating all light nuclei with mass number $A\leq 3$ as active degrees of freedom.
In this setting, deuterons are mainly produced from a proton and a neutron in the reaction $p+n+N\leftrightarrow d+N$, where $N$ is a nucleon acting as a catalyst.
$A=3$ nuclei are produced in a similar way in $4\leftrightarrow 2$ reactions from their three constituents and a catalyst.
The $3\leftrightarrow 2$ and the $4\leftrightarrow 2$ interactions are performed using the stochastic collision criterion~\cite{Staudenmaier:2021lrg}.
As the stochastic collision criterion requires a sufficiently large number of test particles, we represent each particle with 25 test particles  
This method is not available for producing larger nuclei which is important close to the projectile and target rapidities. Therefore, we focus mainly on the results from coalescence but present elliptic flow calculations at midrapidity using the stochastic production of light nuclei.

Compared to the previous study~\cite{Mohs:2020awg}, where the influence of light nuclei formation on flow observables was studied, the dynamic formation of nuclei now includes $A=3$ nuclei and they are treated with the stochastic collision criterion.
The coalescence approach applied in~\cite{Mohs:2020awg} was very simple and used only to distinguish between bound and free nucleons.
The coalescence afterburner applied in the current work can be applied to explicitly form nuclei, allowing to include their contribution in the analysis.


\section{Analysis details}

\subsection{Observables and kinematic variables}

Pressure gradients in the initial stage of the collisions lead to collective expansion of the compressed and heated matter. 
At energies discussed in this work, the presence of the spectator part of the colliding nuclei plays an important role in the final particle anisotropies observed in non-central collisions.
The resulting spatial (early) anisotropy translates into the azimuthal anisotropy of the final-state particles, which are usually quantified by the coefficients in a Fourier expansion of the azimuthal distribution of these particles
\begin{equation}
    \frac{{\rm d}N}{{\rm d} \varphi} \propto 1 + 2 \sum_{n=1}^{\infty} v_{n} \cos \ (n\varphi),
\end{equation}
where $\varphi$ is measured with respect to the reaction plane.
The Fourier coefficients \vdir and \vel are measures of the directed and elliptical flow, respectively. 
These can be calculated from the azimuthal distribution as follows 
\begin{equation}
    v_1 = \langle \cos\varphi \rangle, \quad v_2 = \langle \cos (2\varphi) \rangle.
\end{equation}

Following published data, the results are presented as a function of c.m.s.\ transverse momentum per nucleon and rapidity, both normalized to the projectile values in c.m.s. 
These are defined as  $\pt = (p_{\mathrm{T}}/A)/(p_\mathrm{P}^{\mathrm{c.m.}}/A_{\mathrm{P}})$ and \rap~=~$(y/y_\mathrm{P})^\mathrm{c.m.s.}$, respectively. 
The subscript P denotes characteristics of the projectile. 
In the following, this quantities are called for simplicity transverse momentum and rapidity, although they are normalized quantities.

The impact of the detector acceptance loss for very forward angles ($\theta_{\rm lab}<1.2^\circ$)  was tested and found negligible. 
Thus, this geometrical cut is not applied in the simulations. 

\subsection{Centrality selection}

\begin{figure}[ht!]
    \centering
    \includegraphics[width=0.48\textwidth]{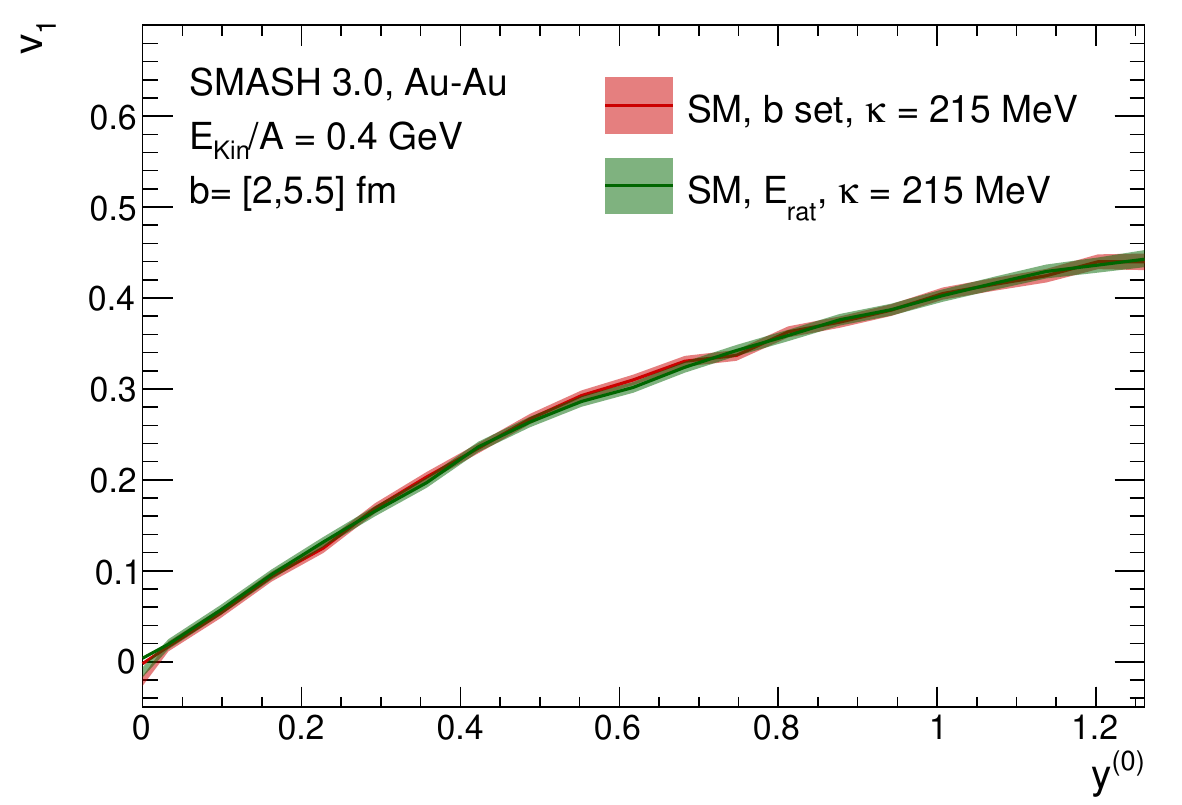}
     \includegraphics[width=0.48\textwidth]{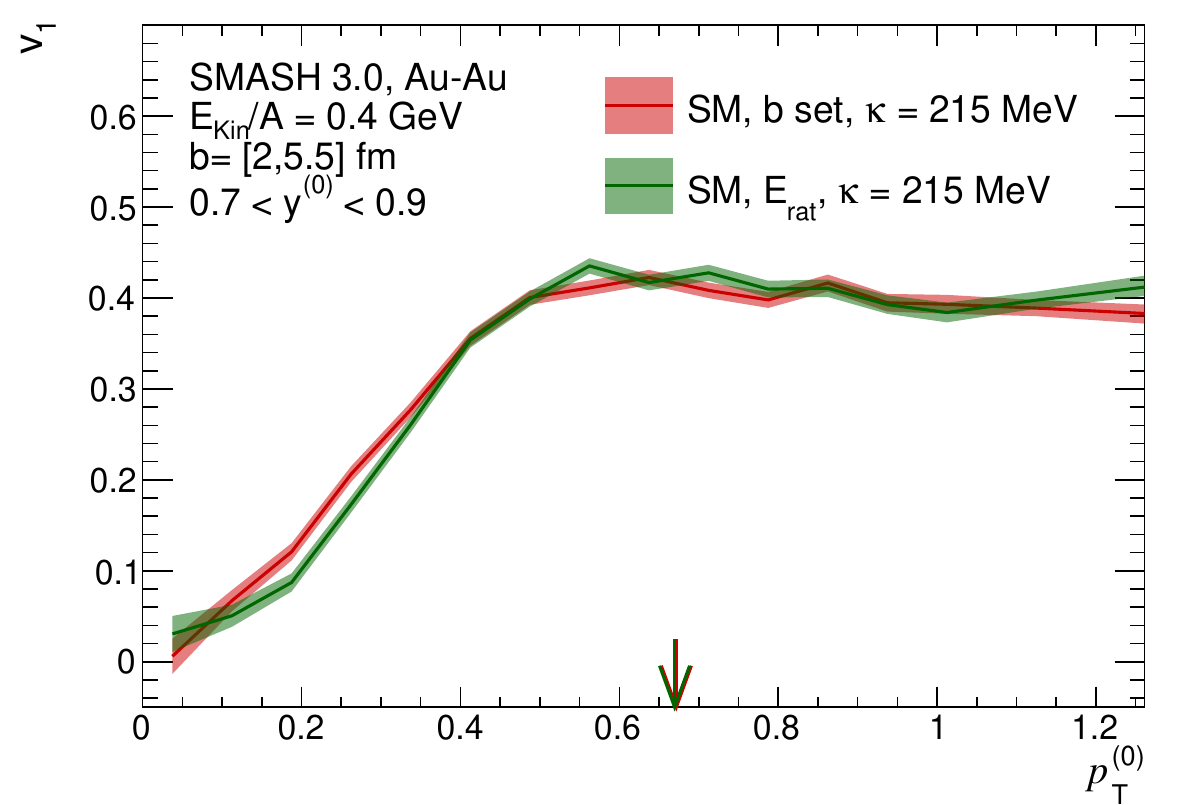}
    \caption{Directed flow coefficient \vdir as a function of rapidity (left) and transverse momentum for \rap~=~0.7 - 0.9 (right) for Au--Au collisions at 0.4~\agev. Two centrality selection methods are compared, based on $b$ and as in data, employing \erat. The~(overlapping) arrows in the right panel indicate the mean \pt.}
    \label{fig:erat_check}
\end{figure}

The default way for the collision centrality selection in the SMASH simulations is a direct setting of the impact parameter $b$.
Since $b$ can not be experimentally measured directly, the~collisions in the~experimental data are divided based on the~charged-particle multiplicity and the~\erat variable defined as
\begin{equation}
    E_{\rm rat} = \frac{\sum_i E_{\perp,i}}{\sum_i E_{\mid\mid,i}} \: , 
\end{equation}
where the~sums run over the~transverse and longitudinal c.m.s.\ kinetic energy components of all the particles detected in an~event. 
First, the multiplicity distribution is divided into classes, based on percentiles of the~inelastic (geometric) cross section. 
Then, based on the correlation between \erat and the multiplicity, the collisions with the highest multiplicities and highest $E_{\rm rat}$ are selected as the most central collisions. 
For the analysis performed here, the actual ranges of $b$ used for different centrality classes are taken from  the assignments performed in Ref.~\cite{FOPI:2004bfz,FOPI:2003fyz}.

In order to check that centrality selections based on the~impact parameter lead to the~same centrality ranges as in the~experimental selection, the~\erat method is used in simulations and compared with the~selection based on the~impact parameter. 
The result of this check is shown in Fig.~\ref{fig:erat_check} for semi-central Au--Au collisions at \ekin=~0.4\,GeV.
As can be seen, the two methods give practically identical results for \vdir, both as a function of rapidity and transverse momentum. 
In the appendix, similar figures are provided also for the other two studied collision systems, Xe--CsI and Ni--Ni, showing consistency of the two methods in these cases, too. 
Based on this, the selection on the impact parameter is used in the following analysis.

\subsection{Particle selection}
\label{sec:part_selection}
Particle selection in the model is performed as for the data, namely $Z=1$ particles.
In SMASH, deuterons and tritons are produced with an afterburner based on the coalescence model as described in Sec.~\ref{sec:lnf}.
In the case of the~study of \vel at midrapidity, the stochastic collision criterion is used as well, as described in Sec.~\ref{sec:lnf}.
In this case, the~simulations are marked in the legends of the figures with an additional "s", e.g.\ HMs stands for Hard EoS with momentum-dependent potentials and the stochastic collision criterion.
The impact of spectator nuclei, which can significantly influence directed flow at forward rapidities, is suppressed by selecting only particles which interacted at least once during the~system evolution. 
This is not possible in case of the~stochastic collision criterion, where also the~particles in the~spectator part of the colliding nuclei interact. 
As a consequence, the stochastic collision criterion is used only for the studies of \vel at midrapidity, where the impact of the spectators is negligible. 

\section{Results}

\begin{figure}[ht!]
    \centering
    \includegraphics[width=0.48\textwidth]{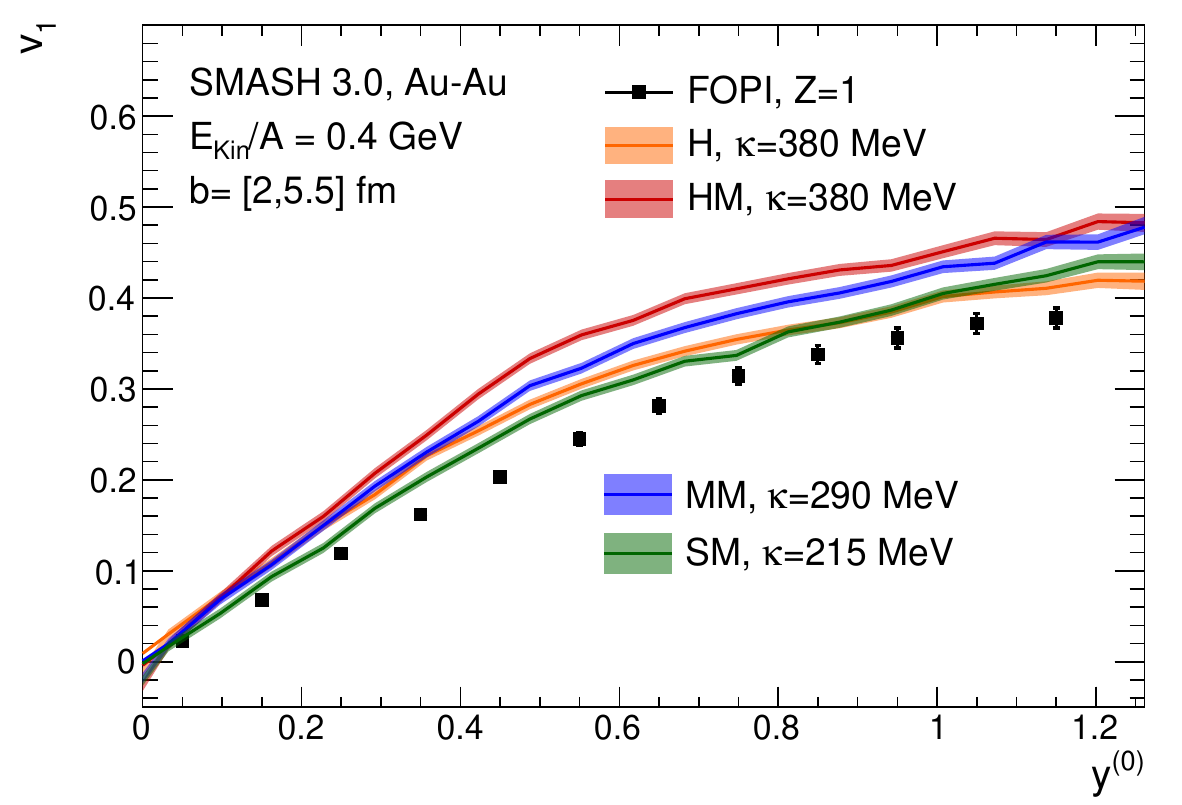}
     \includegraphics[width=0.48\textwidth]{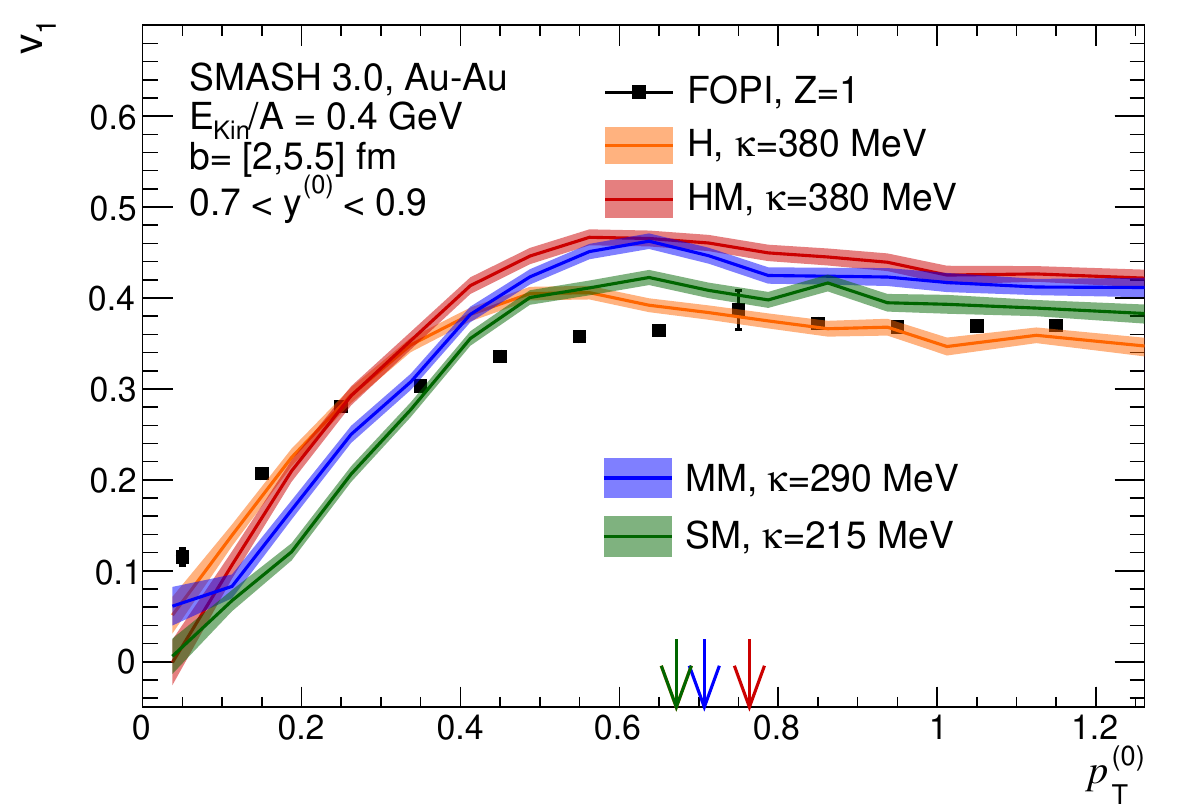}
    \caption{Directed flow coefficient \vdir for different EoS parametrizations and potentials as a function of rapidity (left) and transverse momentum (right) compared with the FOPI data~\cite{FOPI:2003fyz} in Au--Au collisions at \ekin~=~0.4~GeV. The arrows in the right panel indicate \mpt values (note that the orange and dark green arrows overlap).}
    \label{fig:v1_AuAu}
\end{figure}

To study the EoS and the impact of different potentials, we start by investigating the directed flow \vdir of charged particles as a function of rapidity and transverse momentum.
Figure~\ref{fig:v1_AuAu} compares \vdir for Au--Au collisions at \ekin~=~0.4~GeV with a~broad set of model settings.  
It is found that the momentum dependent potentials lead to a higher directed flow for \pt-integrated values (l.h.s.). 
When studying \vdir as a function of \pt in a selected bin of rapidity (r.h.s) it is observed that the larger \vdir values originate from high transverse momenta, while at low \pt there is very little, if any, sensitivity to the momentum-dependent potentials. 
The mean \pt, \mpt, also increases in case of momentum dependent potentials, as visible in the right panel of Fig.~\ref{fig:v1_AuAu}.
The sensitivity on the EoS ($\kappa$) is larger at lower \pt values.
For other collision systems and energies discussed below, only Hard EoS and Soft EoS, both with momentum dependent potentials, are studied as bracketing cases. 

\begin{figure}[h!]
    \centering
    \includegraphics[width=0.48\textwidth]{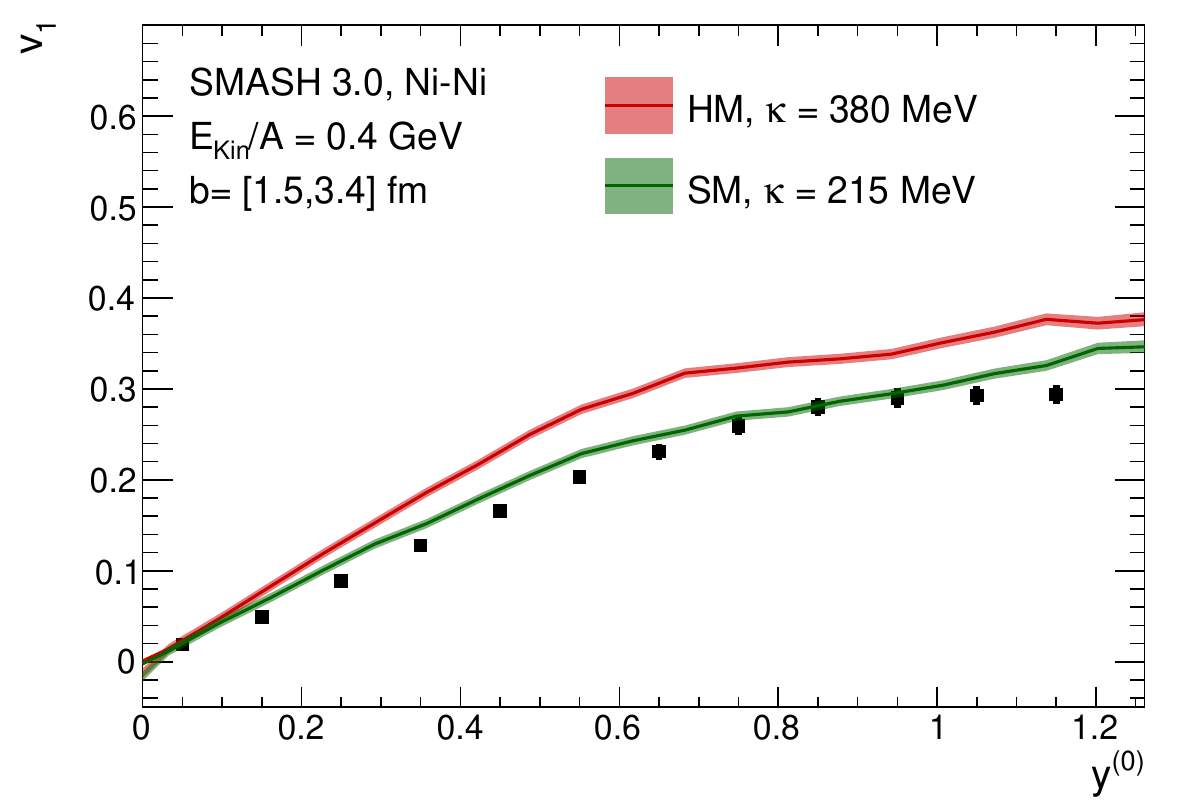}
     \includegraphics[width=0.48\textwidth]{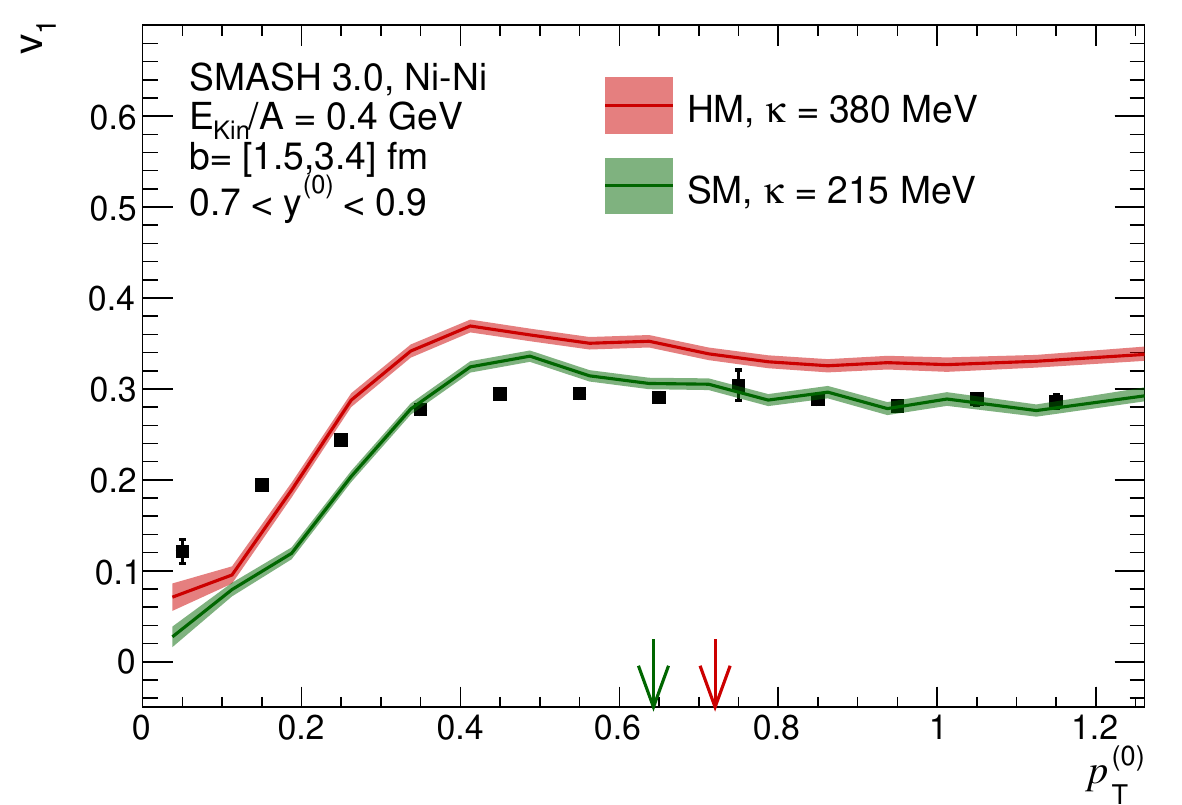}
    \caption{Directed flow coefficient \vdir for Hard and Soft EoS with momentum dependent potential as a function of rapidity (left) and transverse momentum (right) compared with the FOPI data~\cite{FOPI:2003fyz} in Ni--Ni collisions at \ekin~=~0.4 GeV. The arrows in the right panel indicate \mpt values.}
    \label{fig:v1_NiNi}
\end{figure}

\begin{figure}[h!]
    \centering
    \includegraphics[width=0.48\textwidth]{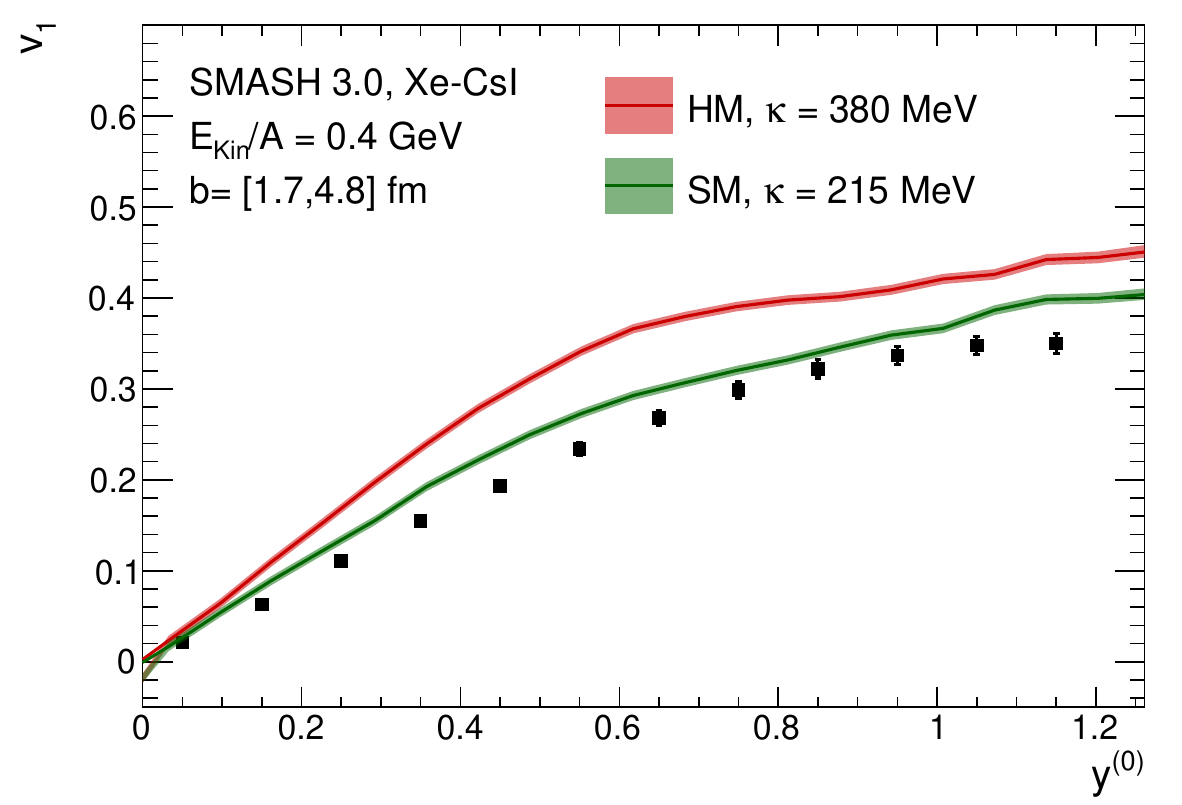}
     \includegraphics[width=0.48\textwidth]{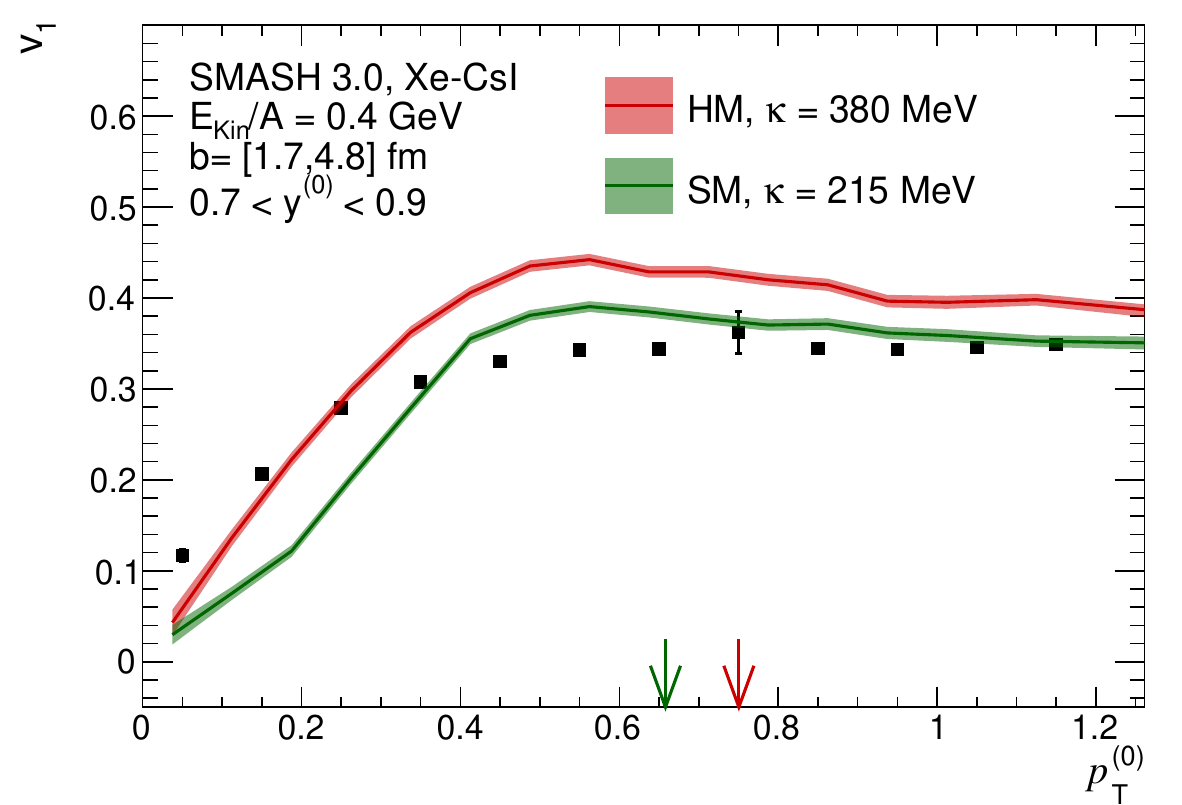}
    \caption{As Fig.~\ref{fig:v1_NiNi} but for Xe--CsI collisions.}
    \label{fig:v1_XeCsI}
\end{figure}

Figures~\ref{fig:v1_NiNi} and \ref{fig:v1_XeCsI} present the results of the directed flow coefficient for Ni--Ni and Xe--CsI collisions, respectively. 
Like in Au-Au collisions, the soft EoS with momentum dependence (SM) reproduces the overall data better than the hard EoS with momentum dependence (HM) within the SMASH model. 
However, for all studied collision systems, one can notice a poor description of the directed flow \vdir at low \pt, which for the SM EoS is significantly weaker than in experimental data. 

The description of directed flow as function of rapidity is in good agreement with the measurements for the soft momentum-dependent EoS. This is due to the fact, that the directed flow close to the mean transverse momentum is in good agreement with the data. At smaller momenta there might be ambiguities in the spectator selection that affect the results.

\begin{figure}[htb]
    \centering
    \includegraphics[width=0.48\textwidth]{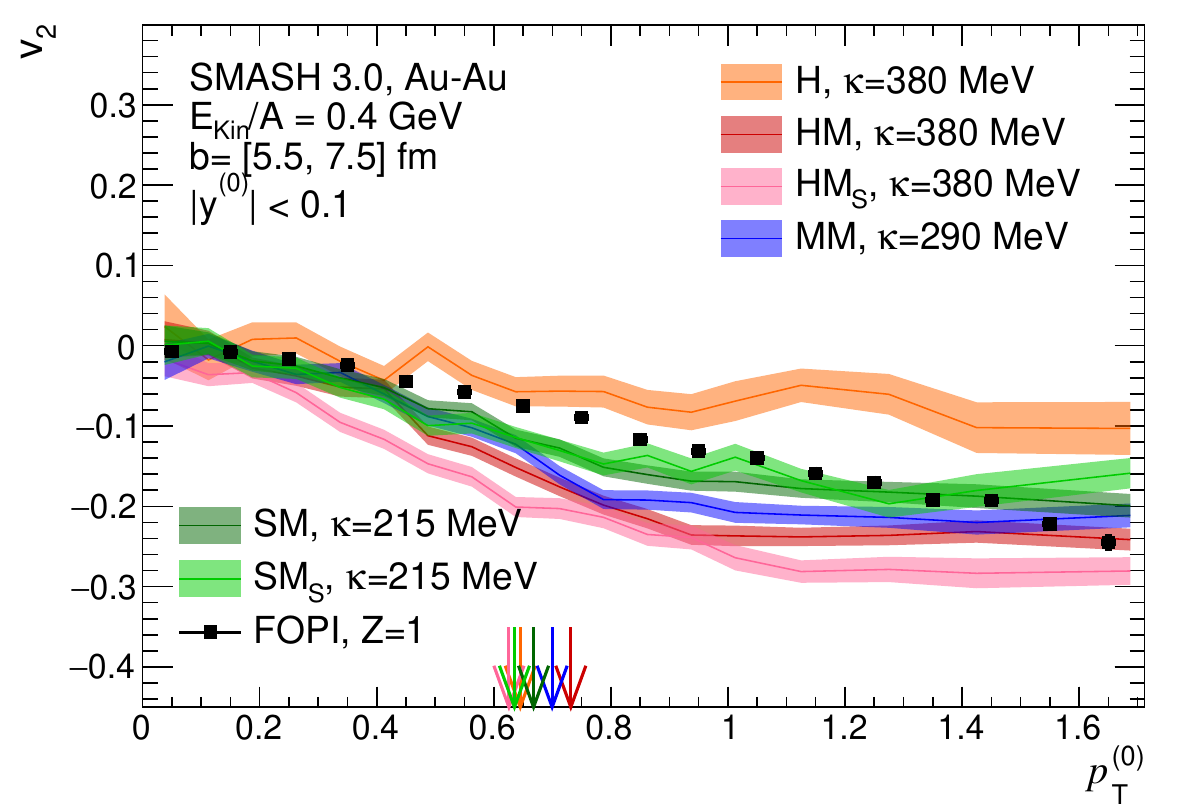}
     \includegraphics[width=0.48\textwidth]{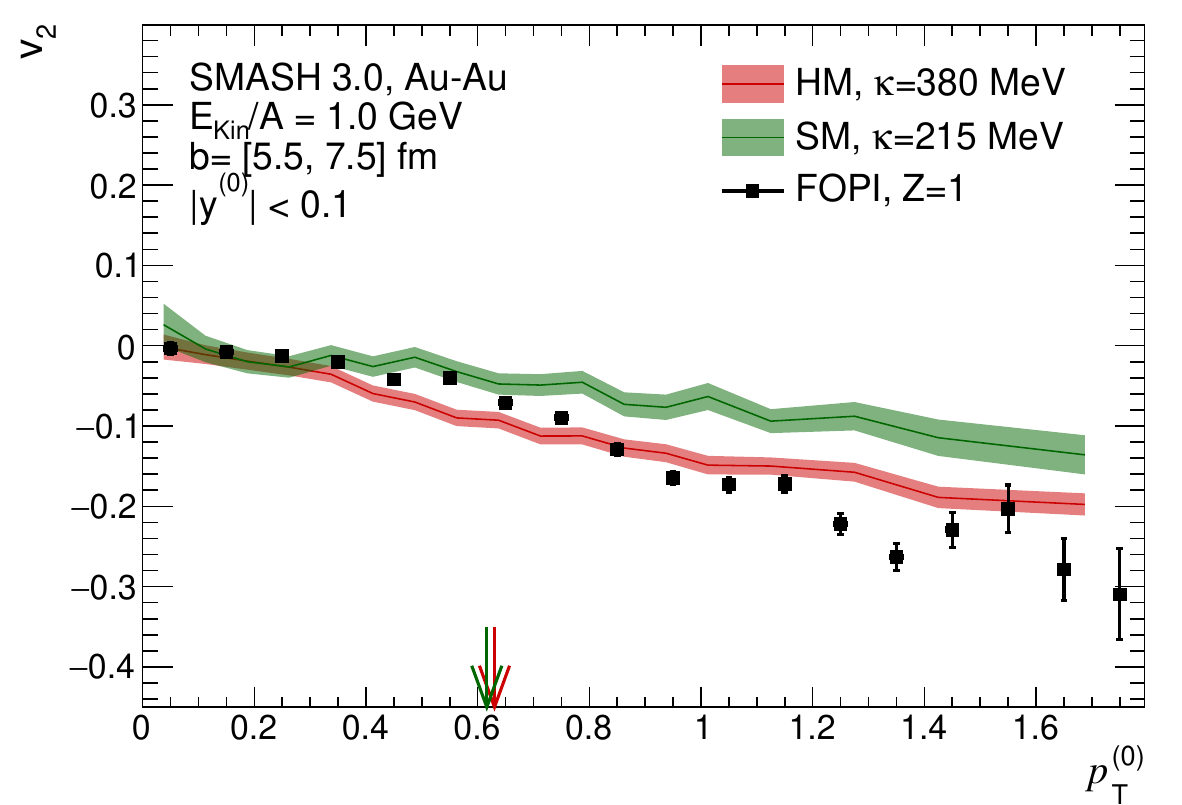}
    \caption{Elliptic flow coefficient \vel for different EoS parametrizations as a function of transverse momentum at \ekin~=~0.4~GeV (left) and \ekin~=~1.0~GeV (right) compared to FOPI data~\cite{FOPI:2004bfz}. HMs stands for Hard EoS with momentum dependent potentials and the stochastic collision criterion. The arrows indicate \mpt values. }
    \label{fig:v2_pt_AuAu}
\end{figure}

The elliptic flow coefficient as a function of \pt in Au--Au collisions at midrapidity is shown in Fig.~\ref{fig:v2_pt_AuAu}. 
Model calculations at \ekin~=~0.4 and 1.0~GeV are compared to FOPI data~\cite{FOPI:2004bfz}.
For the collision energy 0.4~\agev (left panel of Fig.~\ref{fig:v2_pt_AuAu}), the complete set of different EoS versions is shown, including, for some cases, calculations with the~stochastic collision criterion. 
The impact of the momentum-dependent potentials is illustrated and appears significant for the~hard EoS. 
Using a constant potential overshoots the measured elliptic flow coefficients very strongly, while switching on the momentum dependence of the potentials brings the coefficients much closer to the data.
This effect is similar for both the~hard and medium EoS. 
The soft EoS (with momentum dependence) seems to describe the~data on average the~best, but the~trends seen in the data are not fully reproduced by the model. 
A milder dependence on \pt is exhibited by the model starting around \pt = 0.7, which roughly coincides with the mean \pt value. 
At a collision energy of \ekin~=~1~GeV, shown in the right panel of Fig.~\ref{fig:v2_pt_AuAu}, the hard EoS fits the data better, but we notice again that the~\pt dependence of the data is not reproduced well. 
Data-model comparisons for other collision energies between 0.4 and 1.5 \agev are given in Fig. 13 of the Appendix, showing that the~\pt dependence at higher collision energies is fairly well described with hard EoS.
This is in agreement with the results from \cite{Mohs:2024gyc}, where a hard EoS with momentum-dependence is preferred comparing SMASH calculations to flow data at $E_\mathrm{kin}=1.23A\,\mathrm{GeV}$. 

\begin{figure}[htb]
    \centering
    \includegraphics[width=0.48\textwidth]{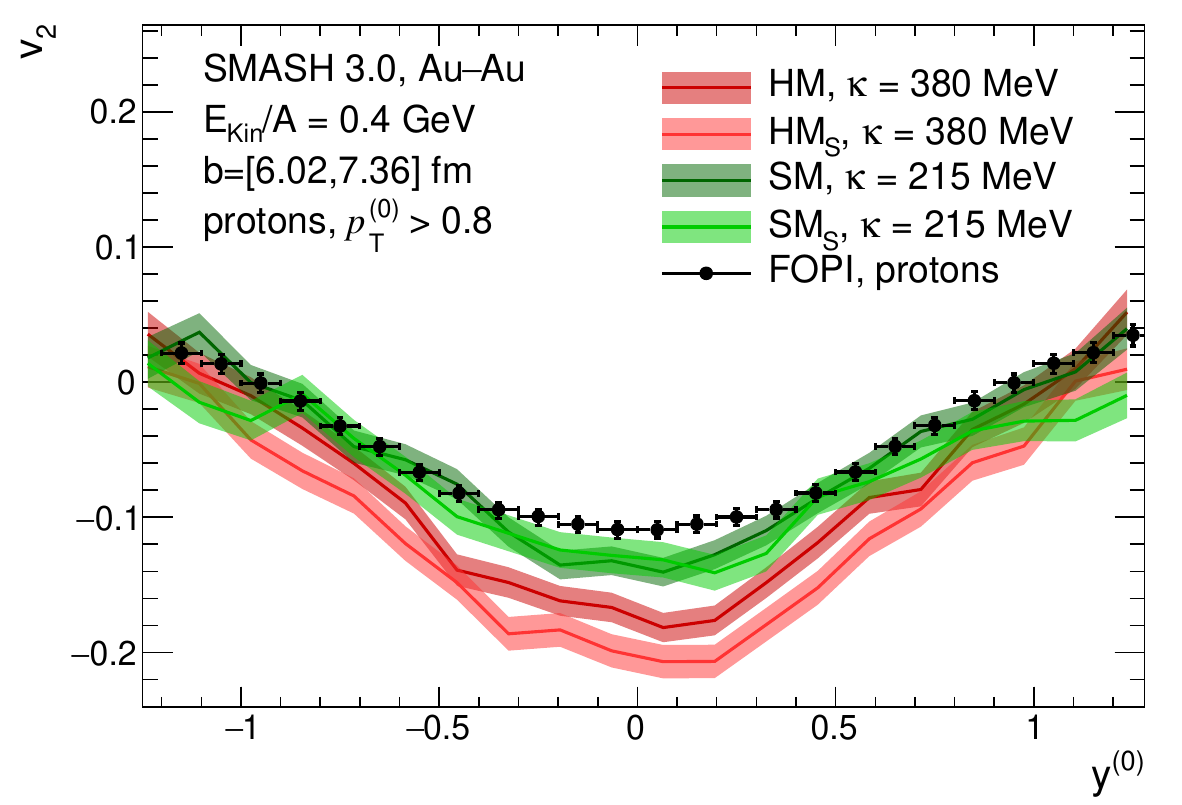}
     \includegraphics[width=0.48\textwidth]{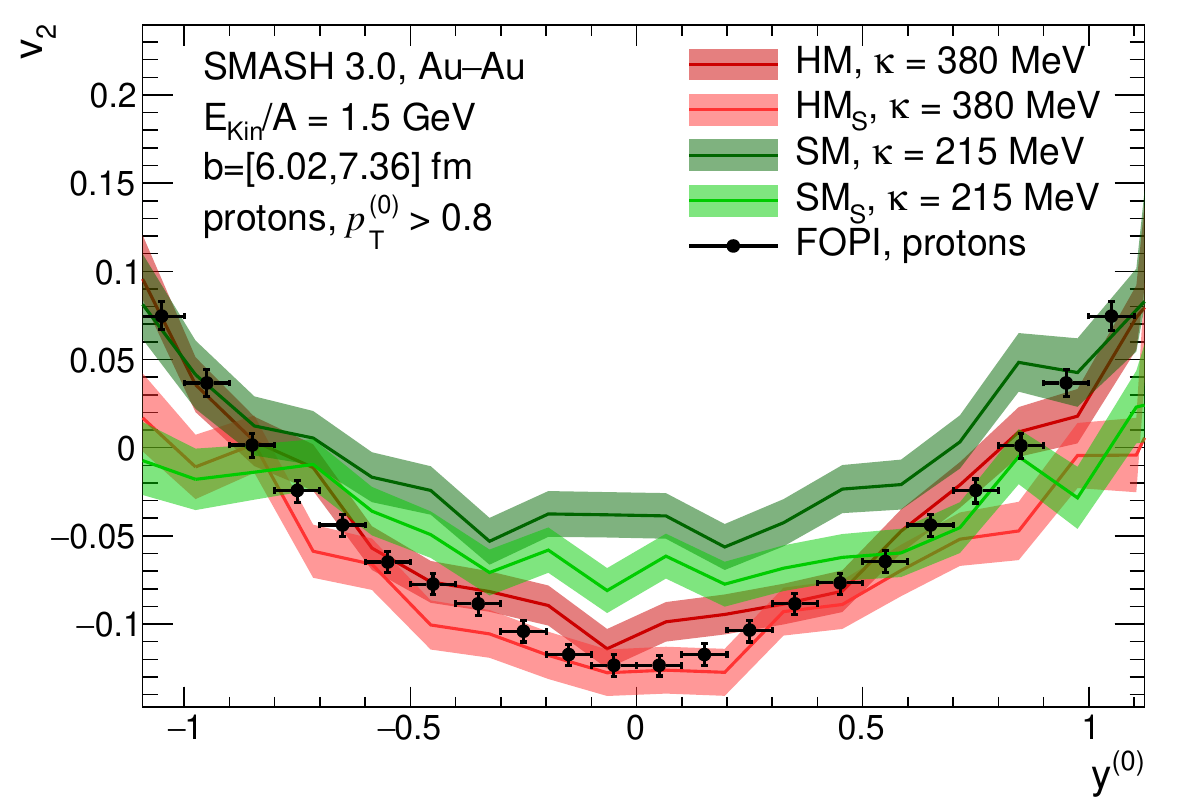}
    \caption{Elliptic flow coefficient \vel of protons, integrated for \pt$>$0.8, as a function of rapidity for mid-peripheral Au--Au collisions at \ekin~=~0.4~GeV (left) and \ekin~=~1.5~GeV (right), compared with FOPI data~\cite{FOPI:2011aa}.
    }
    \label{fig:v2_midPeri_rapidity}
\end{figure}

The elliptic flow coefficient of protons as a function of rapidity in Au--Au collisions at two energies is calculated as well and compared with FOPI data published in~\cite{FOPI:2011aa}. 
For this data set an integration on \pt is performed for \pt~$>$~0.8; note also that the centrality ranges are slightly different compared to the other data sets.
The results for mid-peripheral collisions at \ekin~=~0.4~GeV and \ekin~=~1.5~GeV are shown in Fig.~\ref{fig:v2_midPeri_rapidity}. 
The SMASH predictions with and without the stochastic criterion agree with each other within the rapidity range $|y^{(0)}|<$1. 
It can be observed that at the low collision energy (\ekin~=~0.4~GeV), the soft EoS describes the data clearly better than the hard EoS, while the opposite is true for \ekin~=~1.5~GeV. 
Further comparisons are shown in the Appendix: in Fig.~\ref{fig:v2_midCentral_rapidity}, a similar comparison as in Fig.~\ref{fig:v2_midPeri_rapidity} but for
mid-central collisions, and in Fig.~\ref{fig:v2_ut004_rapidity} a data-to-model comparison of mid-central and mid-peripheral collisions for \pt$>$0.4 at 1.5\agev.

\begin{figure}[t!]
    \centering
    \includegraphics[width=0.48\textwidth]{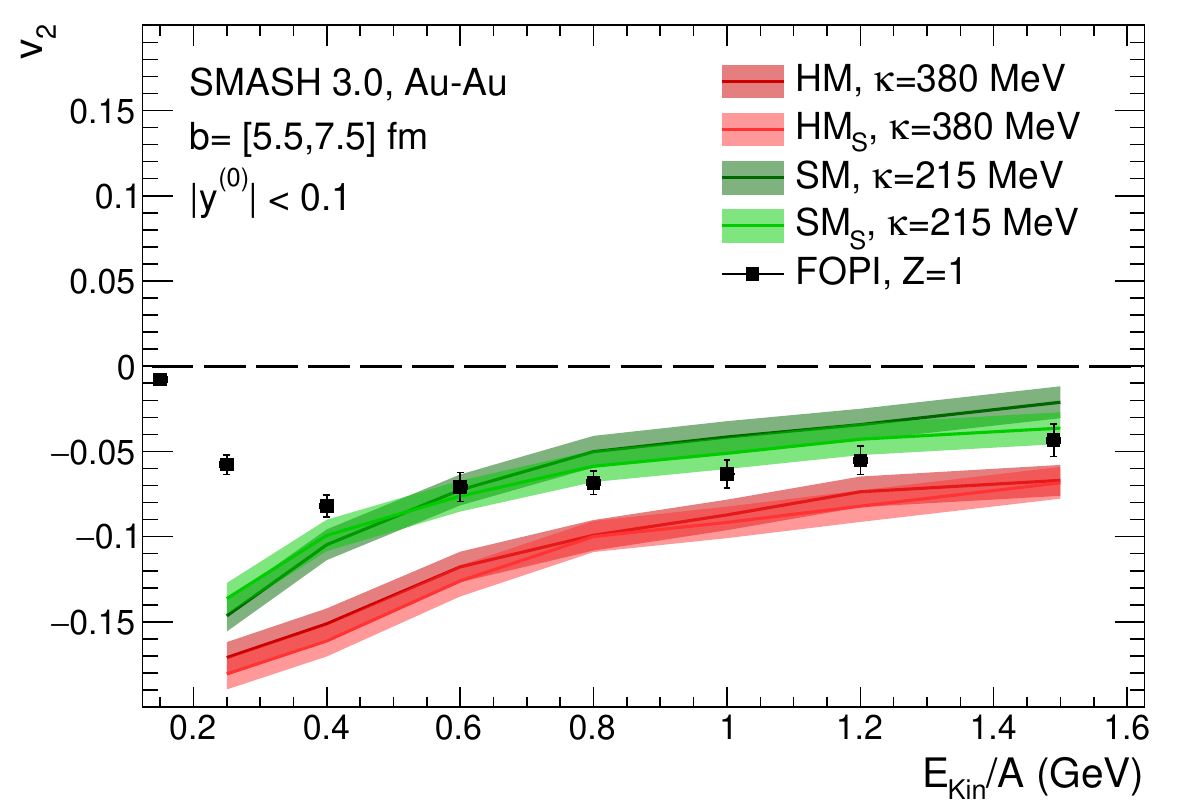}
     \includegraphics[width=0.48\textwidth]{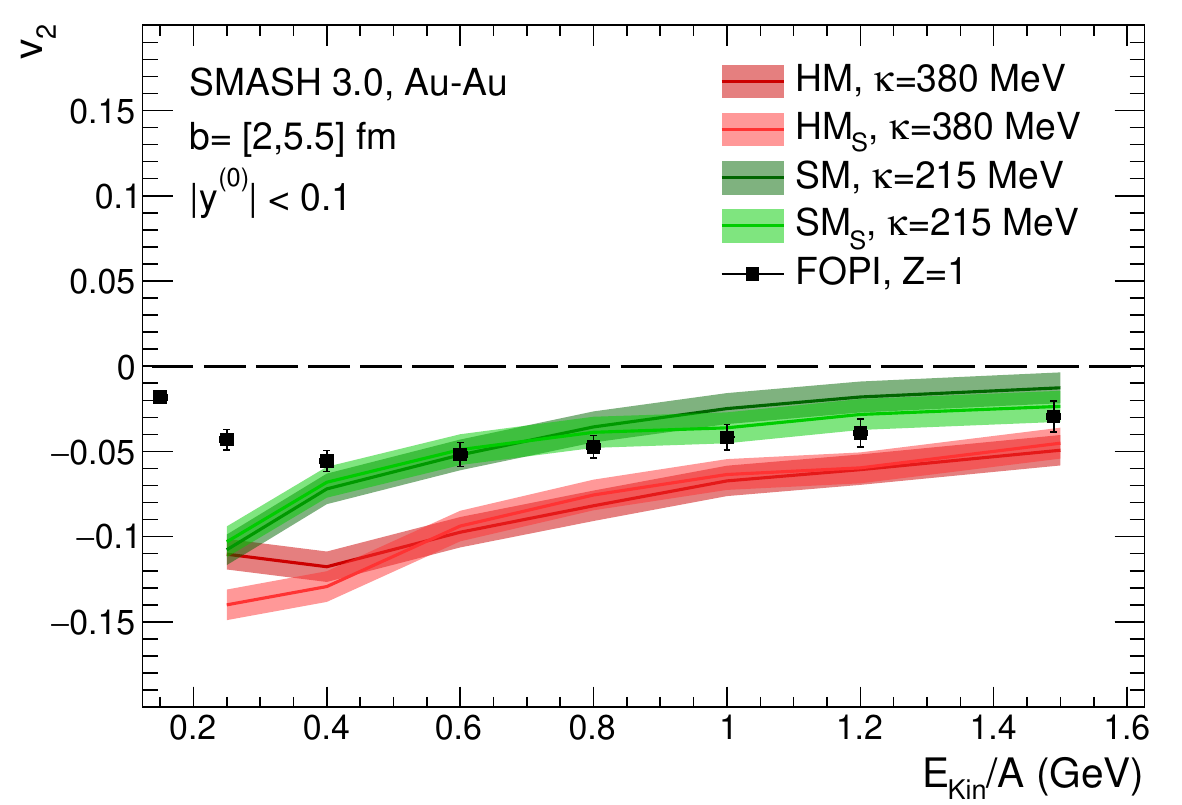}
    \caption{Elliptic flow coefficient of $Z=1$ particles as a function of the beam kinetic energy for two centrality classes, mid-peripheral (left) and mid-central (right), compared with FOPI data~\cite{FOPI:2004bfz}.}
    \label{fig:v2_ekin_AuAu}
\end{figure}

The \pt-integrated elliptic flow at midrapidity in Au--Au collisions as a function of beam energy is presented in Fig.~\ref{fig:v2_ekin_AuAu} for two centrality classes.
As this measurement represents $Z=1$ particles at mid-rapidity, it is not affected by spectators.
Thus, the predictions with and without the stochastic criterion, which are shown as well, largely overlap with one another for both centrality classes and for the different EoSs, in agreement with expectations. 
At the lowest collision energy considered here, \ekin~=~0.25~GeV, the model strongly overestimates the elliptic flow magnitude 
for both EoS parametrizations, suggesting that the model reaches its limits below \ekin~$\simeq$~0.4~GeV. 
At those low beam energies, it is important to correct the cross-sections for the part already treated in potentials by employing medium modified cross-sections, which was not done for the present work.
Generally, both EoS parametrizations, with and without stochastic criterion, predict an energy dependence of the elliptic flow parameter \vel that is significantly stronger than what is found in data. 
Moreover, the hard EoS overestimates the elliptic flow magnitude at all energies and approaches the data only above \ekin~$\simeq$~1.2~GeV.
The soft EoS, on the other hand, describes the data fairly well for collision energies between 0.4 and 0.8~\agev, and slightly underestimates the data at higher energies.
Due to limitations of the model, we disregard the data points at \ekin~=~0.25\,GeV in the $\chi^2/n.d.f.$ calculations presented below and one should keep in mind that conclusions about the EoS are in general based on the model including its shortcomings.

\begin{figure}[t!]
    \centering
    \includegraphics[width=0.48\textwidth]{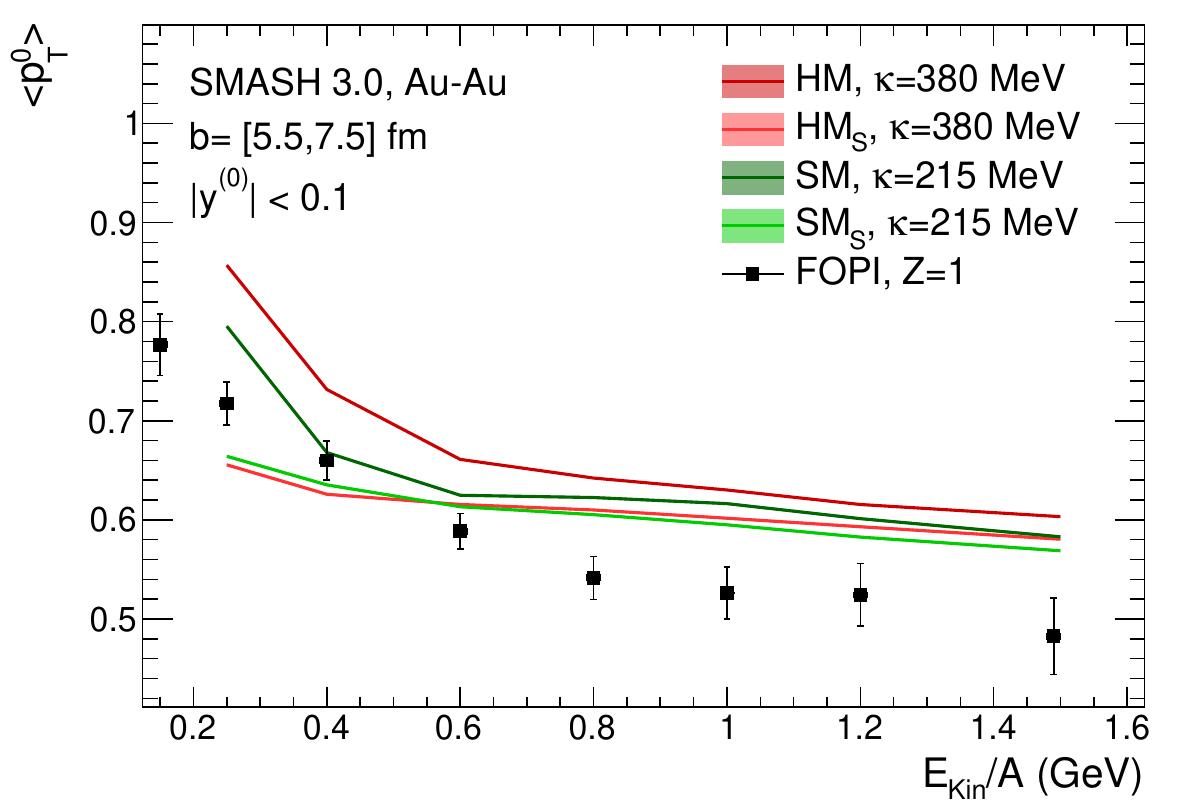}
     \includegraphics[width=0.48\textwidth]{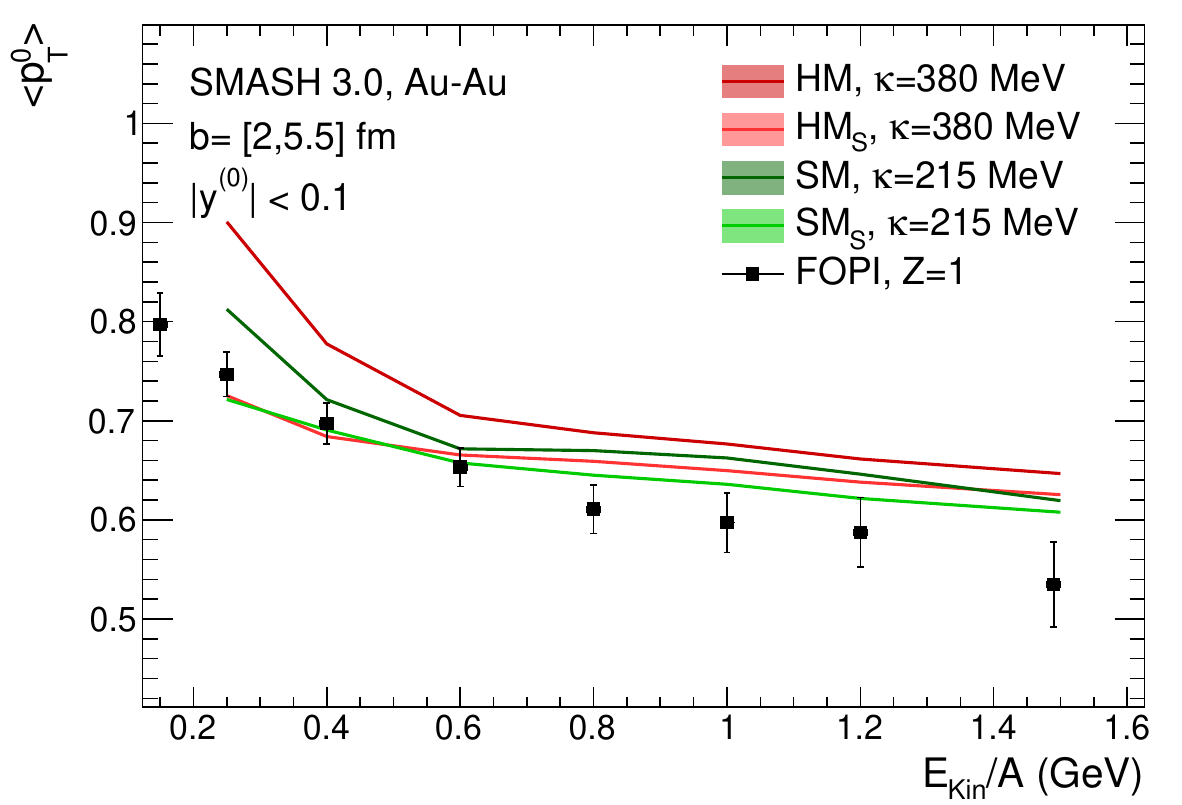}
    \caption{Mean normalized transverse momentum of $Z=1$ particles at midrapidity as a function of the beam kinetic energy for two centrality classes, mid-peripheral (left) and mid-central (right), compared with FOPI data~\cite{FOPI:2004bfz}.
    }
    \label{fig:meanPt_ekin_AuAu}
\end{figure}

The only studied observable for which the calculation with stochastic collision criterion significantly differ from the~geometric one, is the mean transverse momentum for $Z=1$ particles at midrapidity. 
As seen in Fig.~\ref{fig:meanPt_ekin_AuAu}, this difference is significant only for low collision energies while for \ekin~$>$~0.6\,GeV all four sets of simulations lead to rather comparable values for mean \pt. This might be due to the fact that the separation of spectator and participant matter is obscured at lower beam energies and the stochastic collision criterion does not cope well with spectator matter as discussed in Sec.~\ref{sec:part_selection}. 
In the higher energy range, however, the model overestimates the measured mean \pt value for both centrality classes. 

\begin{figure}[t!]
    \centering
\includegraphics[width=0.48\textwidth]{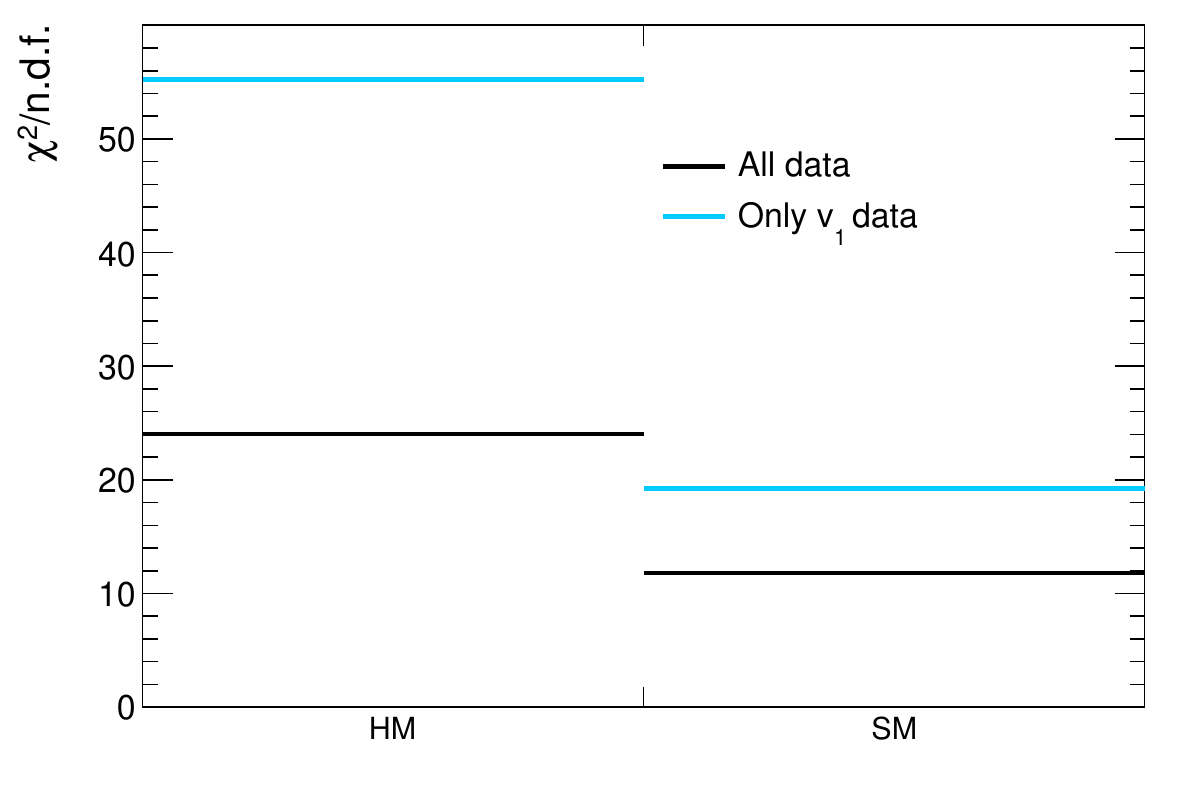}     \includegraphics[width=0.48\textwidth]{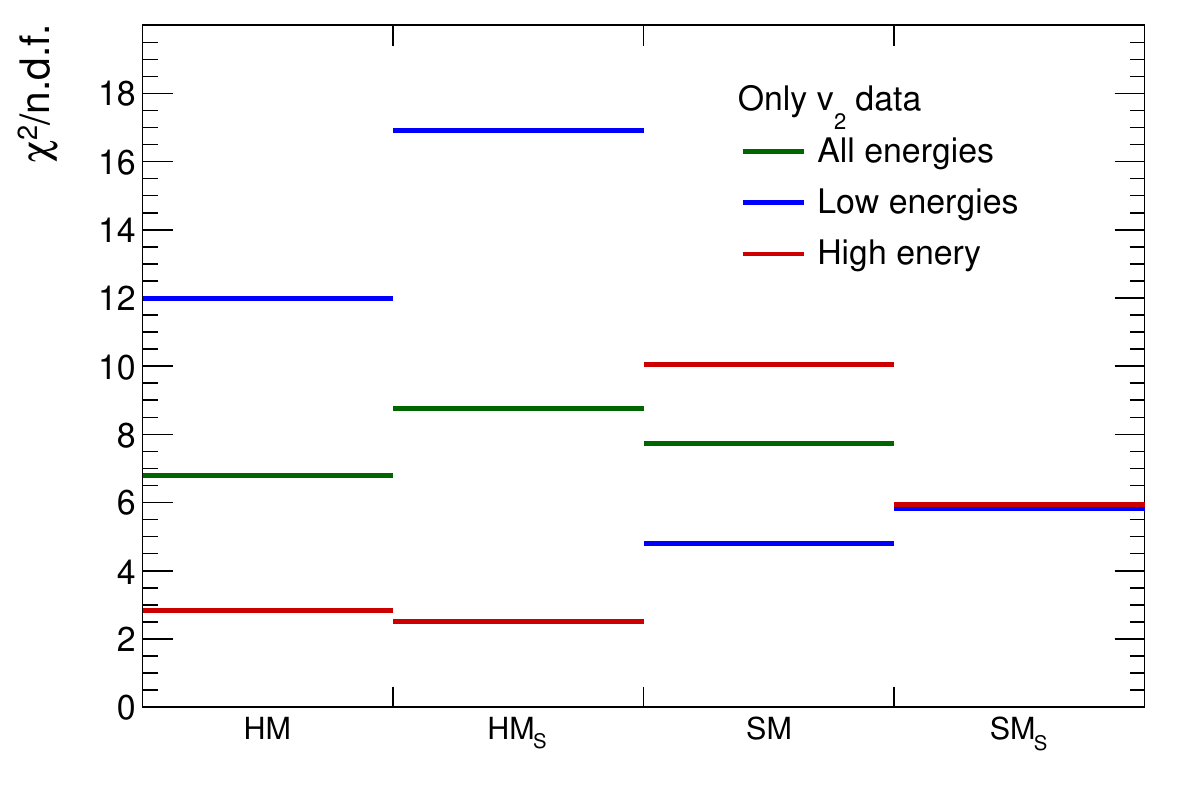}
    \caption{$\chi^{2}$ per number of degrees of freedom between the SMASH predictions for different EoS parametrizations and FOPI data~\cite{FOPI:2004bfz,FOPI:2003fyz}. Left: $\chi^{2}/n.d.f$ using all available data points and only the directed flow data. Right:  $\chi^{2}/n.d.f$ using elliptic flow data, complete and divided in two subsets, for low (0.4, 0.6, 0.8~\agev) and high (1.0, 1.2, 1.5~\agev) collision energies.}
    \label{fig:chi2}
\end{figure}

In order to better quantify which version of EoS (potentials) best describes the experimental data within the model, $\chi^2$ per number of degrees of freedom (\textit{n.d.f.}) is calculated, where the~uncertainties are those of data (statistical and systematic) and model (statistical) added in quadrature. 
The results are summarized in Fig.~\ref{fig:chi2}. 
In the left panel, \chisqrt is shown for all 435 available data points for the standard geometric collision criterion and is compared with \chisqrt for a sub-sample of 155 data points containing only the \vdir results in all collision systems at \ekin~=~0.4~GeV, where the \vdir data points are available. 
In both cases, the soft EoS is characterized by a lower \chisqrt. 
By taking into account only the \vdir data, the \chisqrt for the hard EoS increases by a factor of 2.3 while the one for the soft EoS only by a factor of 1.64, reflecting the large deviations of the HM-EoS from data, seen in Fig.~\ref{fig:v1_AuAu}.
This can be seen also in the right panel of Fig.~\ref{fig:chi2}, where \chisqrt is calculated separately for all the \vel data (279 data points) at low collision energies (\ekin~=0.4, 0.6, and 0.8~GeV, representing a total of 121 data points, including the \pt dependent points in Fig.~\ref{fig:v2_pt_AuAu} and \ref{fig:v2_AuAu_energy}, and rapidity dependent points from Fig.~\ref{fig:v2_midPeri_rapidity},\ref{fig:v2_midCentral_rapidity}, and \ref{fig:v2_ut004_rapidity}), and high collision energies (1.0, 1.2, and 1.5~\agev; 158 data points, including the \pt dependent points in Fig.~\ref{fig:v2_pt_AuAu} and \ref{fig:v2_AuAu_energy}, and rapidity dependent points from Fig.~\ref{fig:v2_midPeri_rapidity},\ref{fig:v2_midCentral_rapidity}, and \ref{fig:v2_ut004_rapidity}). 
The \chisqrt for the soft EoS and stochastic criterion is the same for all three cases, but there is an~ordering for the geometric criterion showing the lowest \chisqrt for lower collision energies. 
This ordering is reversed for the hard EoS for both criteria with larger difference between low and high collision energy.
The comparison of our model to experimental data favors a transition from soft to hard EoS as a~function of energy.
The BUU transport model of Danielewicz~\cite{Danielewicz:1998vz,Danielewicz:2002pu} predicted a similar trend~\cite{FOPI:2004bfz}, while a study within the IQMD model~\cite{LeFevre:2015paj} found a preference for a soft EOS throughout this energy range.
An onset of a softening of EoS is implied by elliptic flow data for \ekin$\simeq\,$2~GeV~\cite{E895:1999ldn,Danielewicz:1998vz}. The conclusions about the stiffness of the EoS in the~energy regime from \ekin=1 GeV and higher also depend on the amount of resonances employed in the transport approach (see e.g.~\cite{Hombach:1998wr}).

\section{Conclusions}

We have performed an exploratory study of the description of directed and elliptic flow in the SMASH transport model at collision energies spanning \ekin~=~0.25 -- 1.5~GeV. While the model clearly shows its limitations at \ekin~=~0.25~GeV, it describes the data well for \ekin$\ge$ 0.4~GeV. 
Clearly, the momentum-dependent potentials are important and the sensitivity to the EoS is significant. 
While overall a soft EoS is preferred by the data, the~elliptic flow data are better described by a hard EoS for the higher collision energies explored here (\ekin=1.0-1.5 GeV). 
Our results are in general consistent with earlier findings on the EoS in other transport approaches.
Further quantitative studies are needed in order to understand systematic uncertainties in transport approaches in general and in SMASH in particular. 
The relevance of better constraints on the EoS from heavy-ion collisions (together with further constraints on the symmetry energy) for neutron stars and their collisions will certainly motivate such studies.

\subsection*{Acknowledgements}
This work has been supported by ``Netzwerke 2021'', an initiative of the Ministry of Culture and Science of the State of North Rhine-Westphalia.
HE and JM acknowledge the support by the State of Hesse within the Research Cluster ELEMENTS (Project ID 500/10.006).
Computational resources have been provided by the HPC cluster PALMA II of the University of Münster, subsidised by the DFG (INST 211/667-1), and by the GreenCube at GSI.

 \bibliographystyle{utphys} 
\bibliography{flow}

\clearpage
\appendix

\begin{figure}[hbt]
    \centering
     \includegraphics[width=0.48\textwidth]{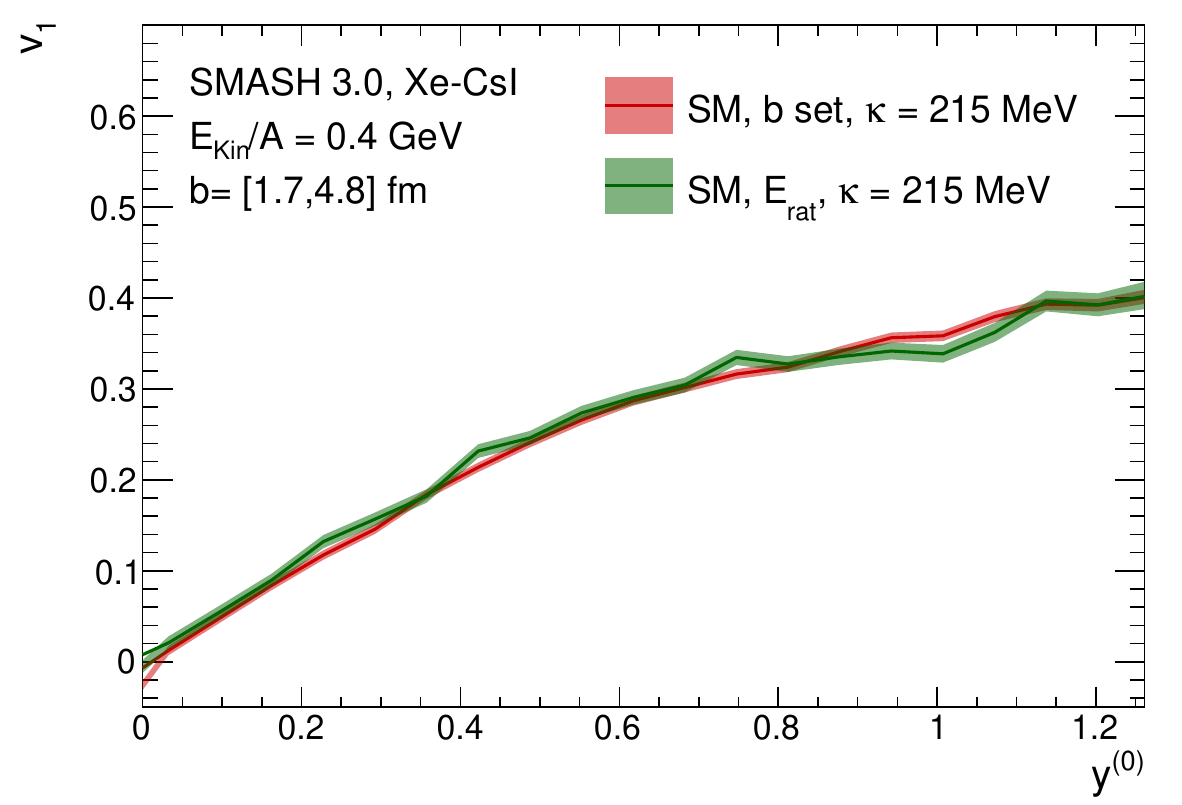}
     \includegraphics[width=0.48\textwidth]{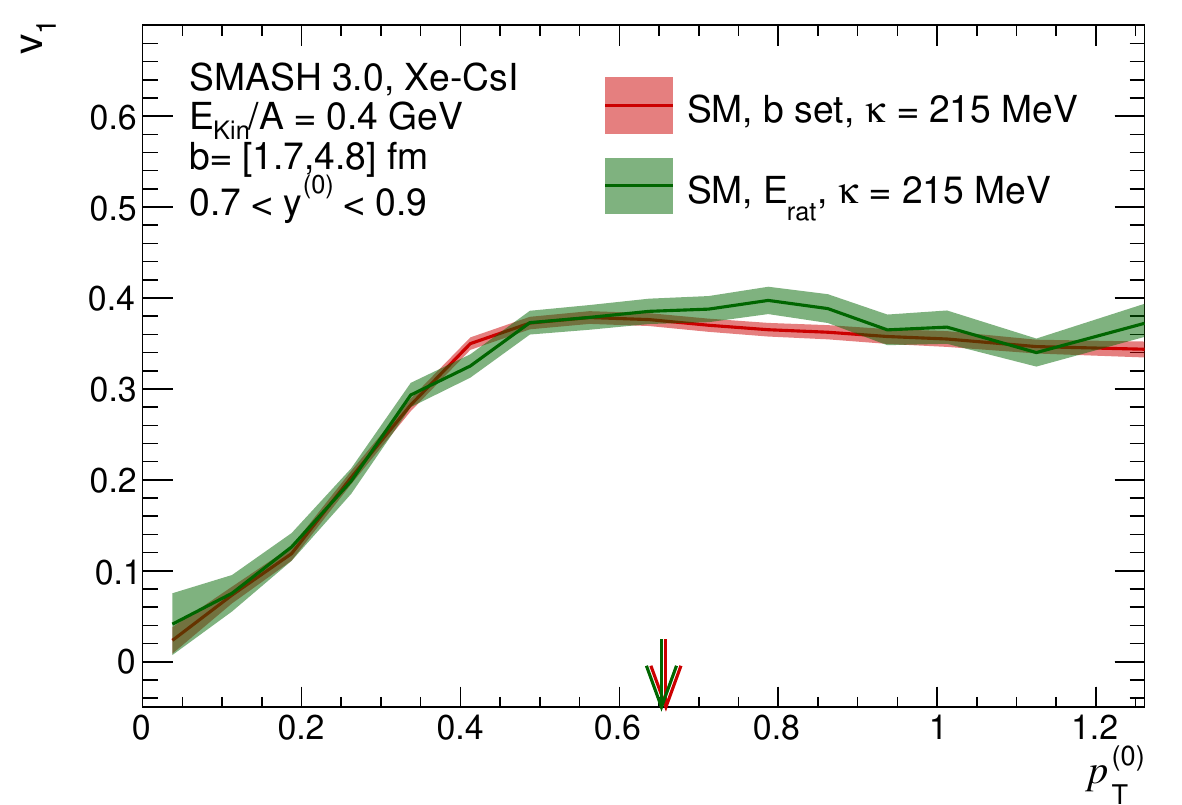}
     \includegraphics[width=0.48\textwidth]{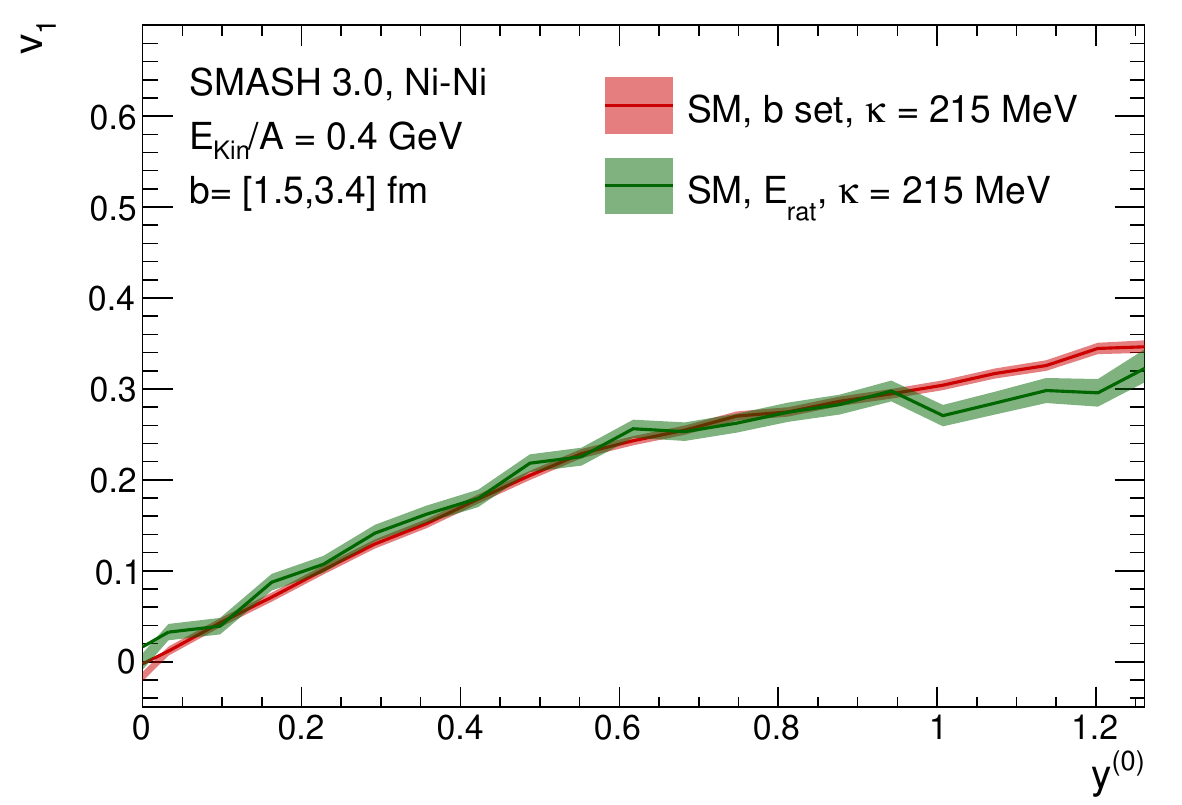}
     \includegraphics[width=0.48\textwidth]{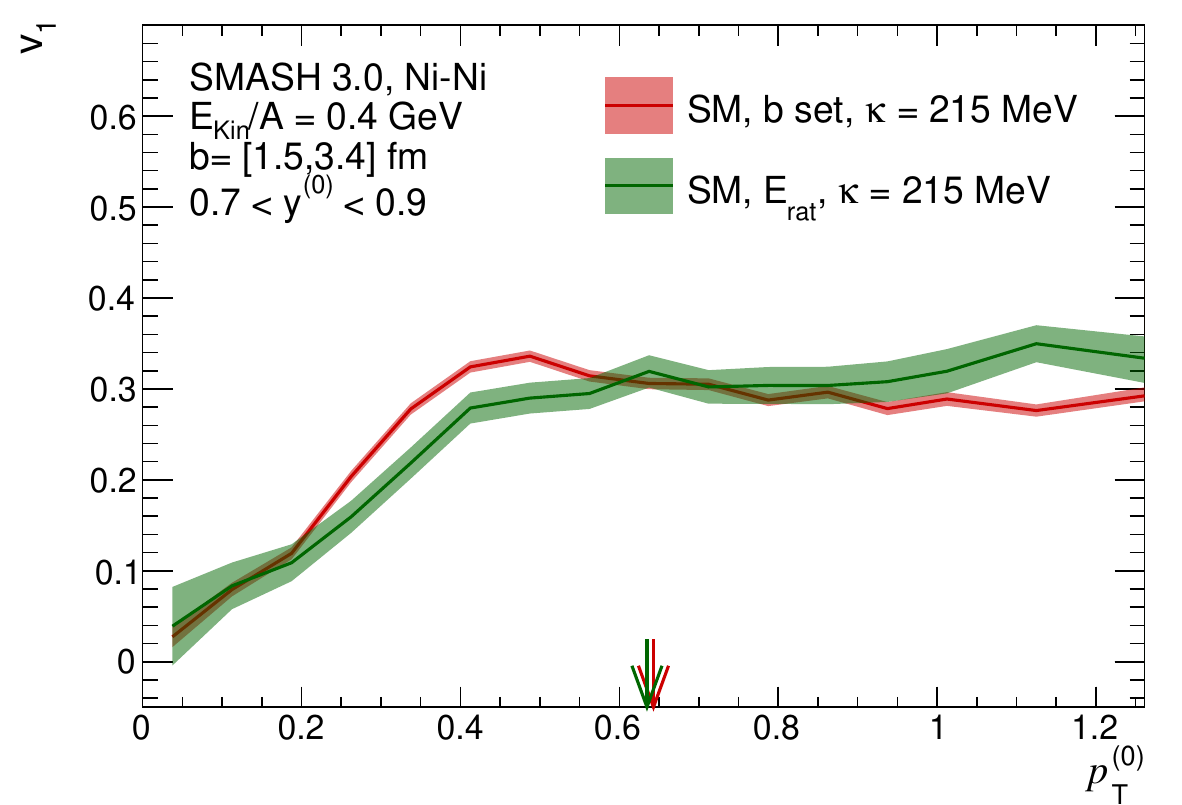}
    \caption{Directed flow coefficient for Xe--CsI (upper panel) and Ni--Ni (lower panel) collisions for centrality selection based on $b$ and \erat.
    }
    \label{fig:erat_check_app}
\end{figure}

\begin{figure}[hbt]
    \centering
    \includegraphics[width=0.48\textwidth]{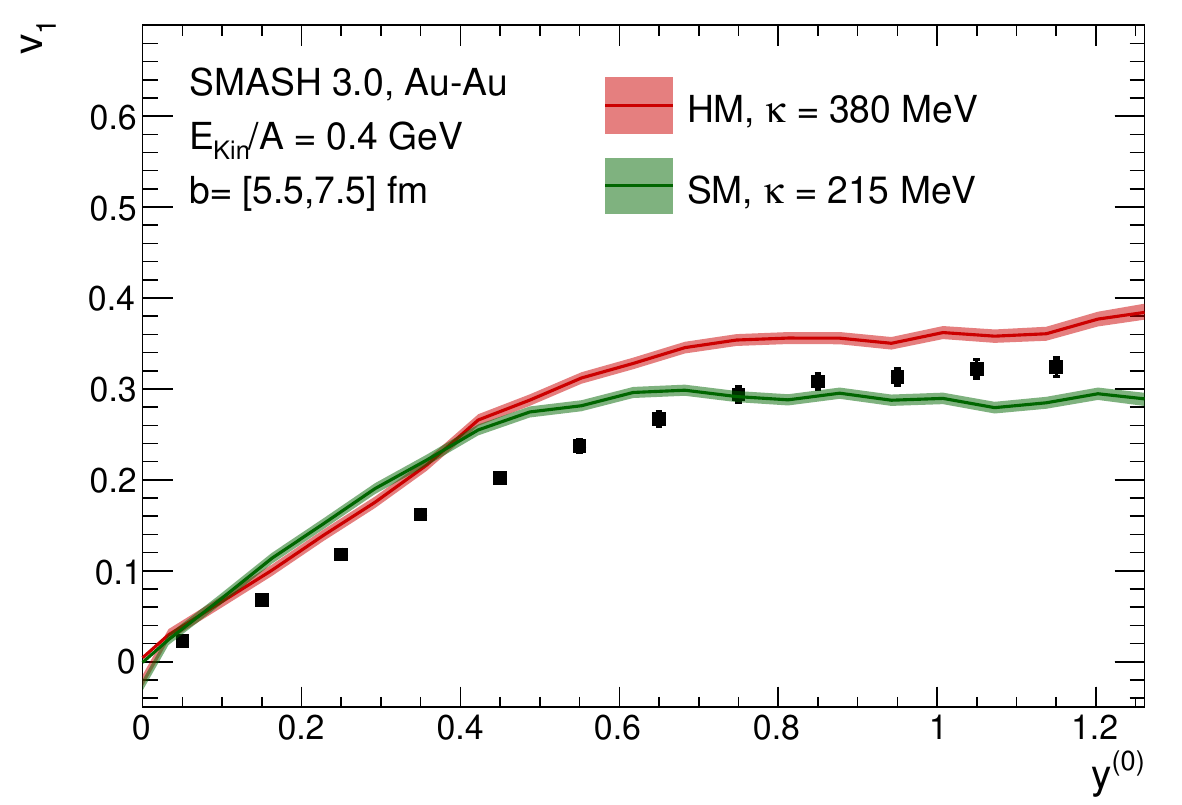}
     \includegraphics[width=0.48\textwidth]{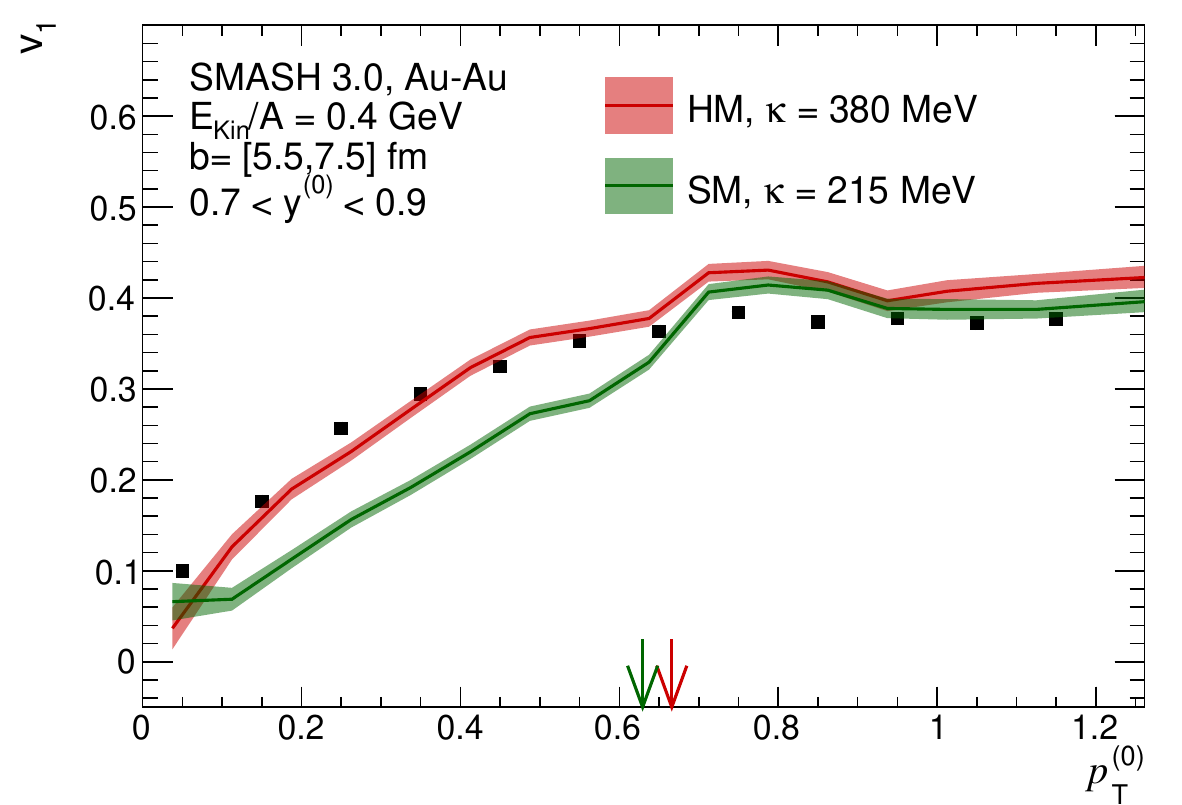}
    \caption{Directed flow coefficient for Au--Au collisions in the centrality class corresponding to $5.5\,{\rm fm} \leq b \leq 7.5\,{\rm fm}$.
    }
    \label{fig:v1_AuAu_app}
\end{figure}

\begin{figure}[hbt]
    \centering
    \includegraphics[width=0.48\textwidth]{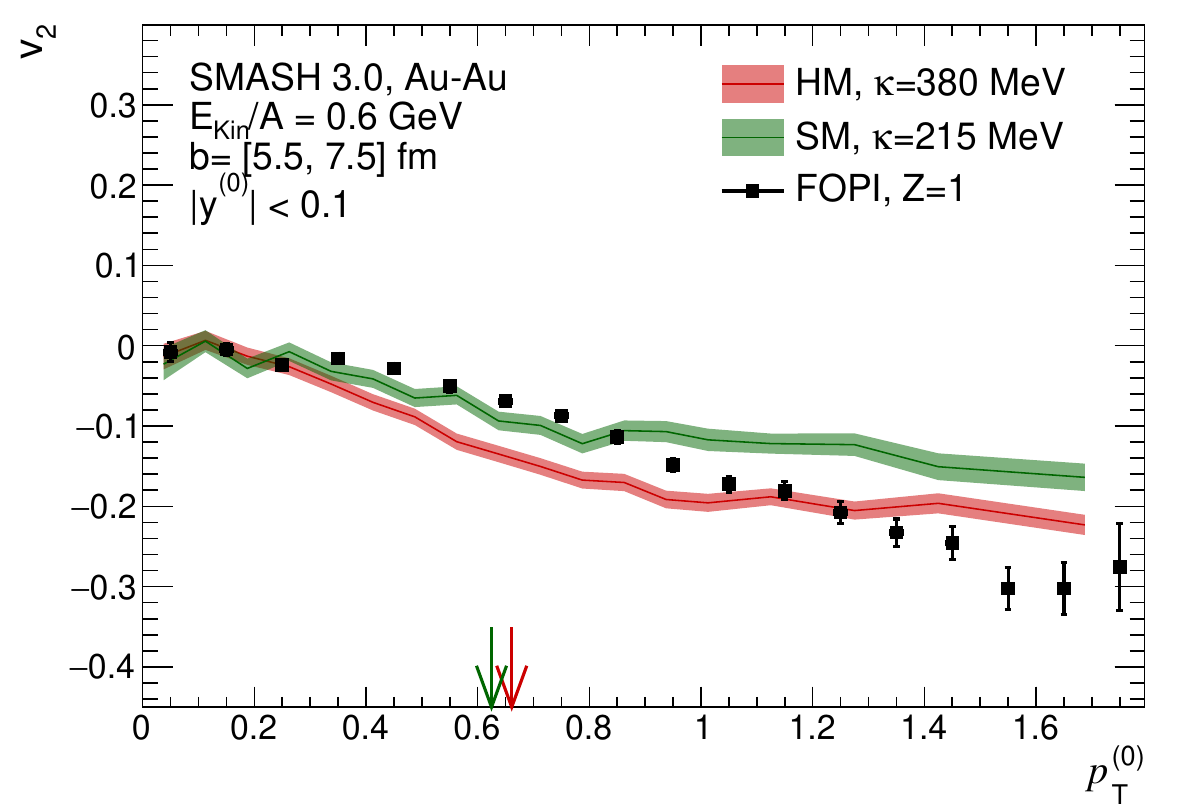}
     \includegraphics[width=0.48\textwidth]{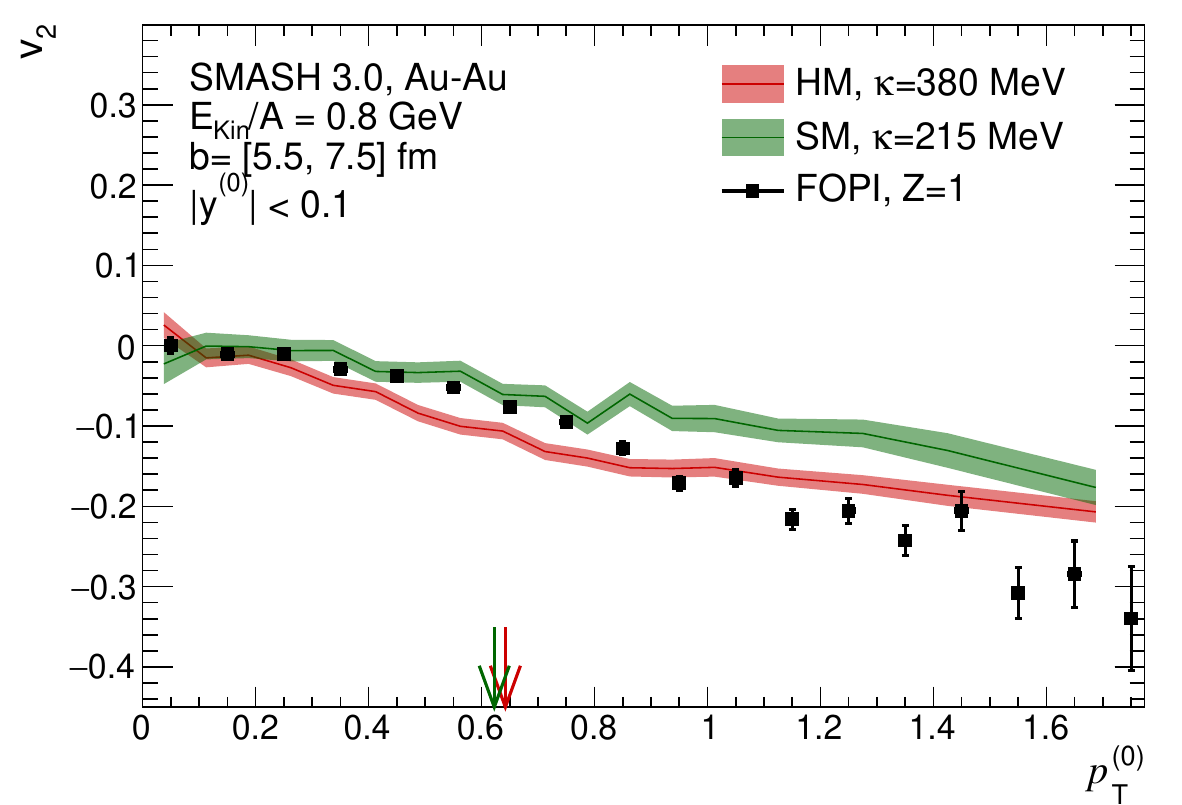}
     \includegraphics[width=0.48\textwidth]{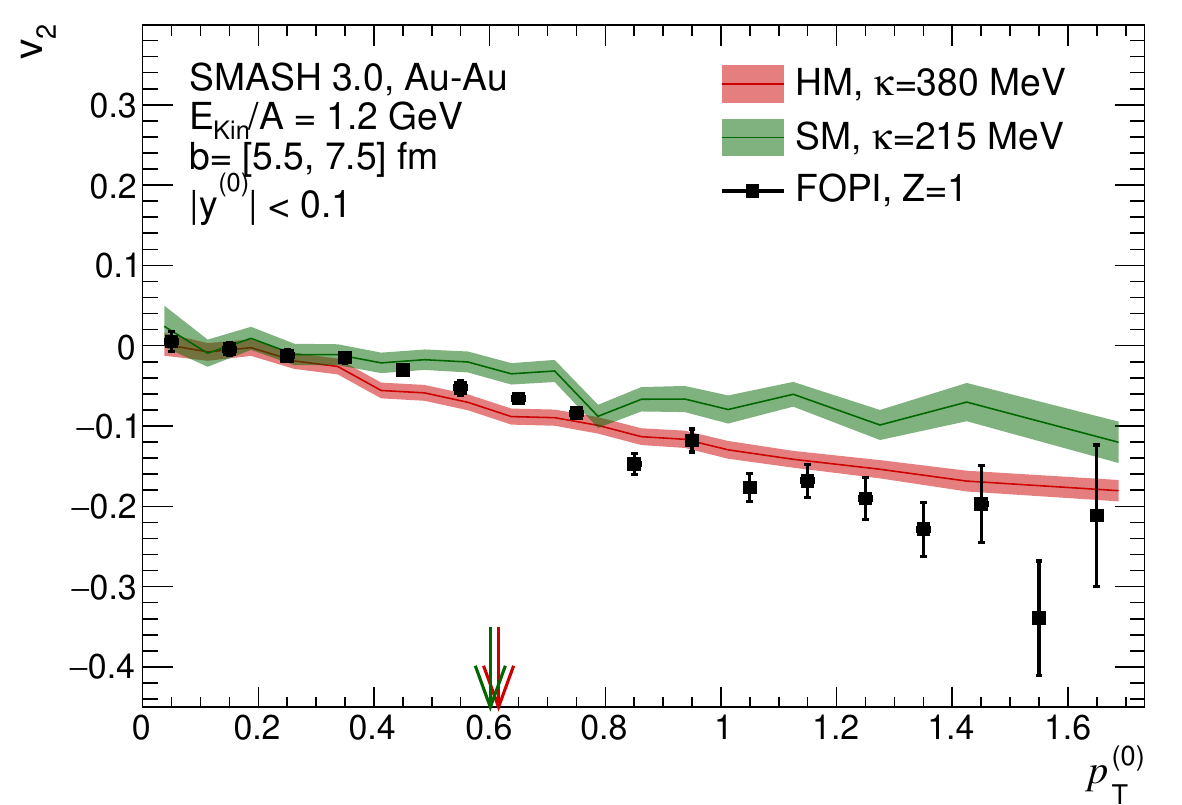}
     \includegraphics[width=0.48\textwidth]{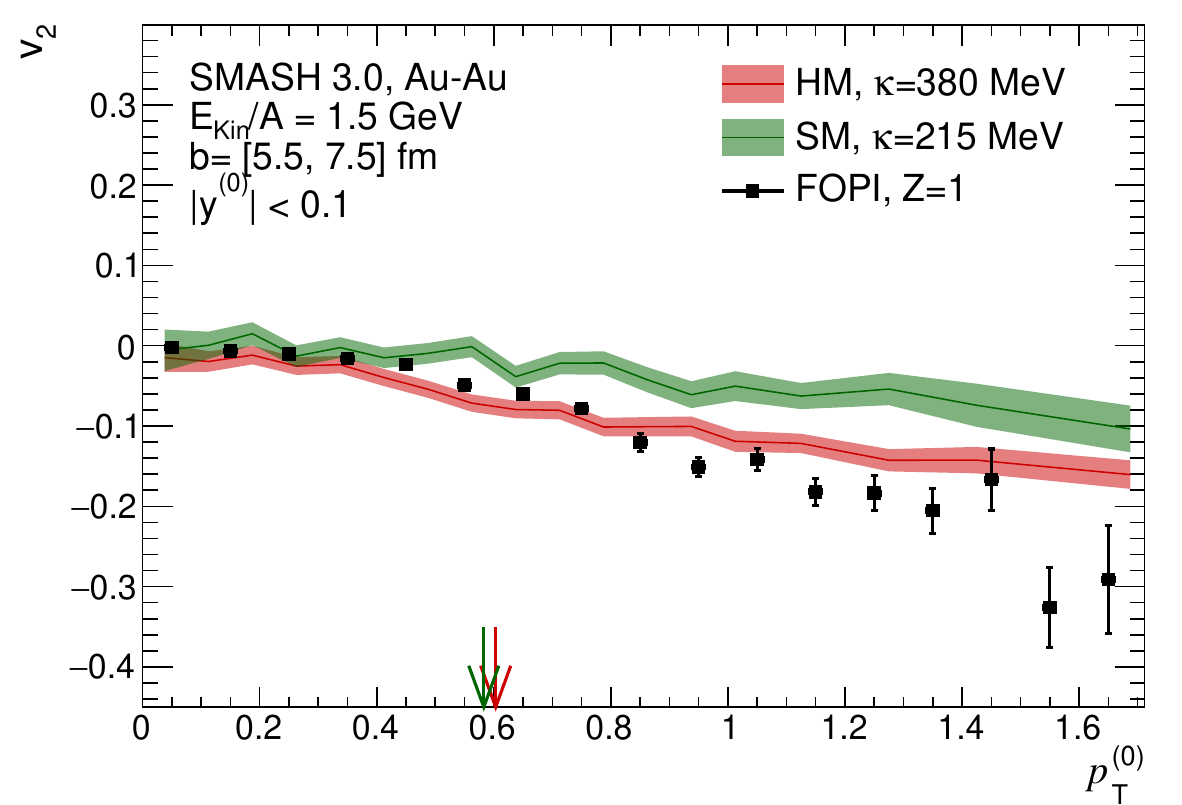}
    \caption{Elliptic flow coefficient as a function of \pt for Au--Au at four collision energies.}
    \label{fig:v2_AuAu_energy}
\end{figure}

\begin{figure}[htb]
    \centering
    \includegraphics[width=0.48\textwidth]{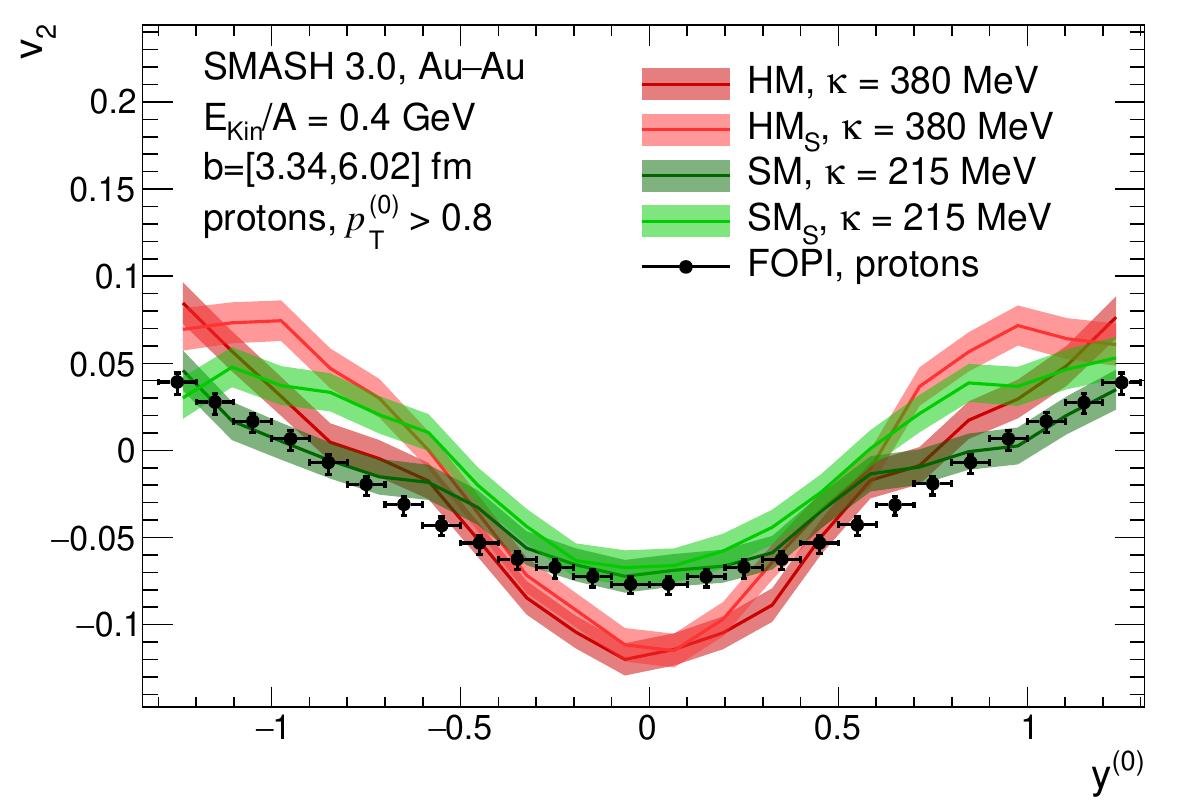}
     \includegraphics[width=0.48\textwidth]{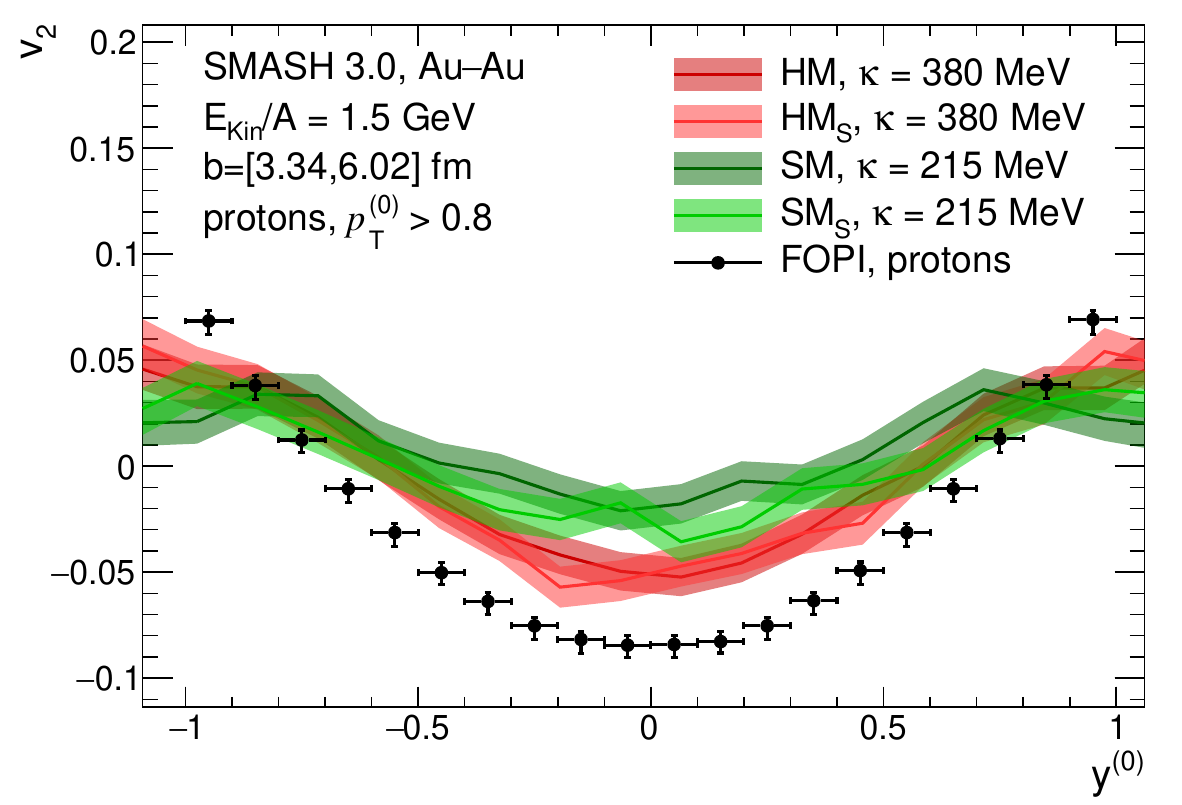}
    \caption{Elliptic flow coefficient \vel for protons with \pt$>$0.8 as a function of rapidity for mid-central Au--Au collisions at \ekin~=~0.4~GeV (left) and \ekin~=~1.5~GeV (right), SMASH calculations compared with FOPI data~\cite{FOPI:2011aa}.
    }
    \label{fig:v2_midCentral_rapidity}
\end{figure}

\begin{figure}[htb]
    \centering
    \includegraphics[width=0.48\textwidth]{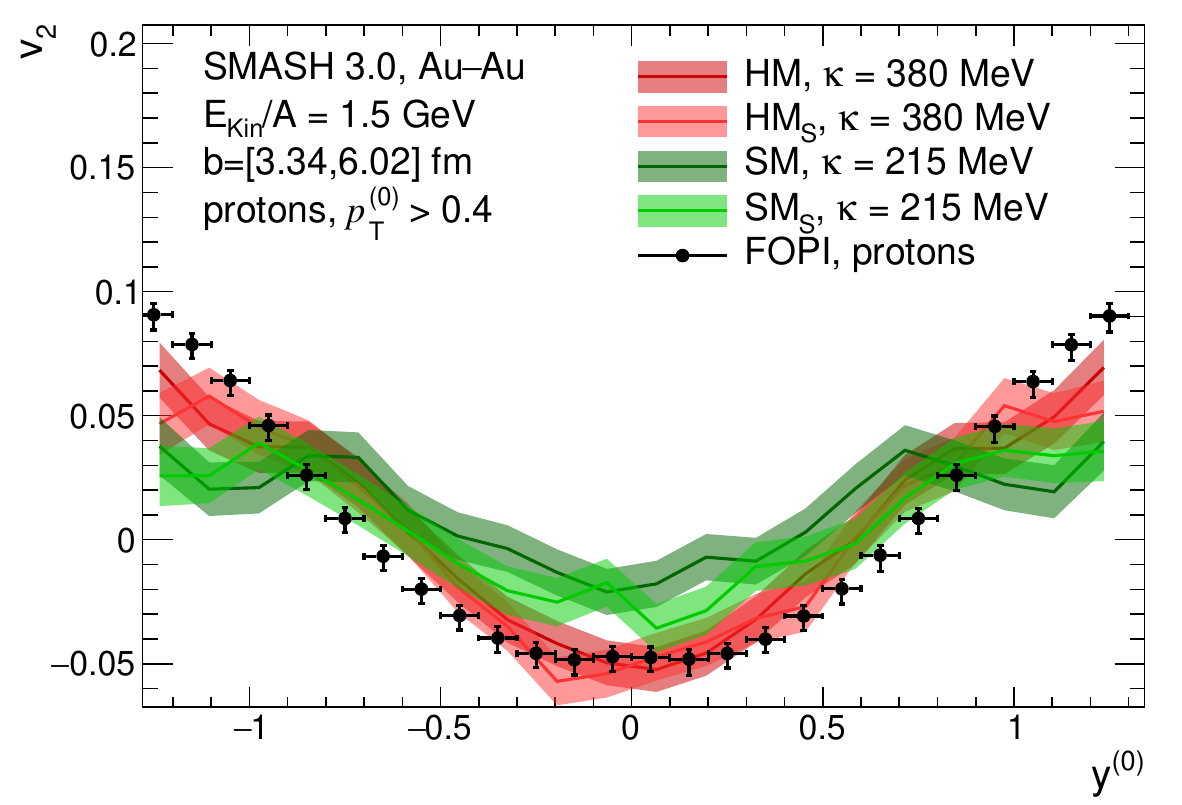}
     \includegraphics[width=0.48\textwidth]{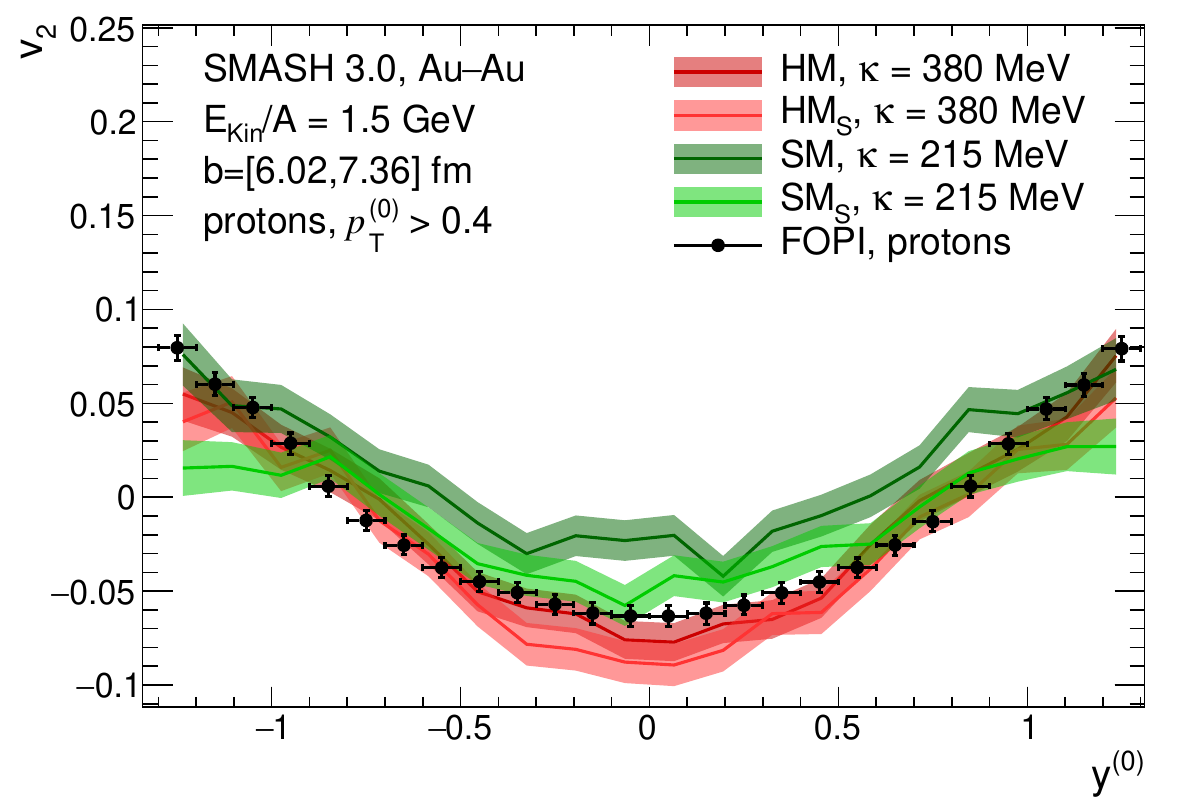}
    \caption{Elliptic flow coefficient \vel for protons with \pt$>$0.4 as a function of rapidity for mid-central (left) and mid-peripheral (right)  Au--Au collisions at \ekin~=~1.5~GeV, SMASH calculations compared with FOPI data~\cite{FOPI:2011aa}.
    }
    \label{fig:v2_ut004_rapidity}
\end{figure}


\end{document}